\documentclass[longauth]{aa} % for the long lists of affiliations 
\usepackage{graphicx}
\usepackage[colorlinks=true,citecolor=blue]{hyperref}
\usepackage{txfonts}
\usepackage[title]{appendix}
\usepackage{xcolor}
\usepackage{siunitx}

\begin{document}

   \title{The {\em Gaia}-ESO survey: mapping the shape and evolution of the radial abundance gradients with open clusters\footnote{Tables A.1, A.2, A.3, A.4, A.5, A.6, A.7, A.8, A.9, A.10, A.11 are only available in electronic form
at the CDS via anonymous ftp to cdsarc.u-strasbg.fr (130.79.128.5)
or via http://cdsweb.u-strasbg.fr/cgi-bin/qcat?J/A+A/}}
%\subtitle{Constraining the mixing process}
\titlerunning{The Galactic abundance gradients} 
\authorrunning{Magrini et al.}
%   \subtitle{}

\author{L. Magrini\inst{\ref{oaa}} \and 
C. Viscasillas V{\'a}zquez\inst{\ref{vilnius}}\and 
L. Spina\inst{\ref{oapd}}\and 
%major comments
S. Randich\inst{\ref{oaa}} \and 
D. Romano\inst{\ref{oabo}}\and 
E. Franciosini\inst{\ref{oaa}} \and
A. Recio-Blanco\inst{\ref{oca}} \and 
T. Nordlander \inst{\ref{canberra}, \ref{astro3d}}\and
V. D'Orazi\inst{\ref{oapd}, \ref{uniroma}}\and 
M. Baratella\inst{\ref{leibniz}}\and
R. Smiljanic\inst{\ref{ncac}} \and 
M.~L.~L.~Dantas \inst{\ref{ncac}} \and
L. Pasquini \inst{\ref{eso}} \and
%minor comments
E. Spitoni\inst{\ref{oca}} \and 
G. Casali\inst{\ref{unibo},\ref{oabo}}\and
M. Van der Swaelmen\inst{\ref{oaa}},
T. Bensby \inst{\ref{lund}} \and 
E. Stonkute \inst{\ref{vilnius}}\and
S. Feltzing \inst{\ref{lund}} \and 
G.~G. Sacco\inst{\ref{oaa}} \and 
A. Bragaglia\inst{\ref{oabo}}\and 
E. Pancino\inst{\ref{oaa}, \ref{ssdc}} \and 
U. Heiter\inst{\ref{uppsala}} \and 
%M. Molero \inst{\ref{units}} \and  
K. Biazzo \inst{\ref{oarm}}\and 
%builders, no commments
G. Gilmore\inst{\ref{cambridge}}\and
M. Bergemann \inst{\ref{maxplank}, \ref{bohr}} \and
G. Tautvai{\v s}ien{\. e}\inst{\ref{vilnius}} \and 
C. Worley\inst{\ref{cambridge}}\and
A. Hourihane\inst{\ref{cambridge}}\and
A. Gonneau\inst{\ref{cambridge}}\and
L. Morbidelli\inst{\ref{oaa}} 
}

\institute{INAF - Osservatorio Astrofisico di Arcetri, Largo E. Fermi 5, 50125, Firenze, Italy \email{laura.magrini@inaf.it} \label{oaa} 
\and
Institute of Theoretical Physics and Astronomy, Vilnius University, Sauletekio av. 3, 10257 Vilnius, Lithuania \label{vilnius} 
\and
INAF - Padova Observatory, Vicolo dell'Osservatorio 5, 35122 Padova, Italy\label{oapd} 
\and
INAF - Osservatorio di Astrofisica e Scienza dello Spazio di Bologna, via Gobetti 93/3, 40129, Bologna, Italy\label{oabo} 
\and
Université Côte d’Azur, Observatoire de la Côte d’Azur, CNRS, Laboratoire Lagrange, Nice, France, \label{oca}
\and
Research School of Astronomy and Astrophysics, Australian National University, Canberra, ACT 2611, Australia\label{canberra}
\and
ARC Centre of Excellence for Astronomy in Three Dimensions (ASTRO-3D), Canberra, ACT 2611, Australia\label{astro3d}
\and
Dipartimento di Fisica, Università degli Studi di Roma Tor Vergata, via della Ricerca scientifica 1, 00133, Roma, Italy\label{uniroma}
\and
Leibniz-Institut f\"ur Astrophysik Potsdam (AIP) An der Sternwarte 16, 14482 Potsdam \label{leibniz}
\and
Nicolaus Copernicus Astronomical Center, Polish Academy of Sciences, ul. Bartycka 18, 00-716, Warsaw, Poland\label{ncac} 
\and
ESO, Karl Schwarzschild Strasse 2, 85748 Garching, Germany \label{eso} 
\and 
Dipartimento di Fisica e Astronomia, Università degli studi di Bologna, via Gobetti 93/2, I-40129 Bologna, Italy \label{unibo}
\and
Lund Observatory, Department of Astronomy and Theoretical Physics, Box 43, SE-221 00 Lund, Sweden \label{lund}
\and
Space Science Data Center - agenzia Spaziale Italiana, via del Politecnico, s.n.c., I-00133, Roma, Italy\label{ssdc}
\and
Observational Astrophysics, Department of Physics and Astronomy, Uppsala University, Box 516, 75120 Uppsala, Sweden\label{uppsala}
\and
INAF - Rome Observatory, Via Frascati, 33, Monte Porzio Catone (RM), Italy\label{oarm} 
\and
Institute of Astronomy, University of Cambridge, Madingley Road, Cambridge CB3 0HA, United Kingdom\label{cambridge}
\and %
Max Planck Institute for Astronomy, Koenigstuhl 17, 69117, Heidelberg, Germany \label{maxplank}
\and
Niels Bohr International Academy, Niels Bohr Institute, University of Copenhagen Blegdamsvej 17, DK-2100 Copenhagen, Denmark \label{bohr}
}

   \date{}

  \abstract
   {The spatial distribution of elemental abundances and their time evolution are among the major constraints to disentangle the scenarios of formation and evolution of the Galaxy.  }
   {In this paper, we used the sample of open clusters available in the final release of the {\em Gaia}-ESO survey to trace the Galactic radial abundance and abundance to iron ratio gradients, and their time evolution. }
   {We selected member stars in 62 open clusters, with ages from 0.1 to about 7~Gyr, located in the Galactic thin disc at Galactocentric radii (R$_{\rm GC}$) from about 6 to 21~kpc. We analysed  the shape of the resulting [Fe/H] gradient, the average gradients [El/H] and [El/Fe] combining elements belonging to four different nucleosynthesis channels, and their individual abundance and abundance ratio gradients. We also investigated the time evolution of the gradients dividing open clusters in three age bins. }
   {The  [Fe/H] gradient has a slope of $-$0.054~dex~kpc$^{-1}$. It can be better approximated with a two-slope shape, steeper for R$_{\rm GC}\leq$11.2~kpc and flatter in the outer regions. We saw different behaviours for elements belonging to different channels. Regarding the time evolution of the gradient, we found that the youngest clusters (age<1~Gyr) in the inner disc have lower metallicity than their older counterpart and they outline a flatter gradient. We considered some possible explanations, including the effects of gas inflow and migration. We suggested that the most likely one may be related to a bias introduced by the standard spectroscopic analysis producing lower metallicities in the analysis of low gravity stars.       }
   {To delineate the shape of the `true' gradient, we most likely should limit our analysis to stars with low surface gravity  $\log g$>2.5 and microturbulent parameter $\xi$<1.8 km~s$^{-1}$. Based on this reduced sample, we can conclude that the gradient has minimally evolved  over the time-frame outlined by the open clusters, indicating a slow and stationary formation of the thin disc in the latest Gyr.  
    We found a secondary role of clusters' migration in shaping the gradient, with a more prominent role of migration for the oldest clusters.    }

   \maketitle
%
%____________________________
%____________________________________
\section{Introduction}
\label{section:introduction}
Open clusters (OCs)
%, thanks to their widespread distribution in the Galactic thin disc and to the secure determination of ages and distances through isochrone fitting of their whole sequence,  
are considered excellent tracers of the chemical properties of the thin disc stellar populations of our Galaxy, including the spatial distribution of elemental abundances.  
Since the end of the '70s, several works exploited the use of OCs to trace the radial distribution, initially, of metallicity, then of individual elements \citep[see, e.g.][]{mayor76, janes79, janes88,  friel93, carraro94, friel95, twarog97, friel02}. 
Considered on their own, or appropriately combined with other stellar and nebular Galactic tracers, which correspond to different epochs in the evolution of our Galaxy, such as for instance,  H~II regions \citep[e.g.][]{peimber78, balser11, esteban17,  arellano20, mendez22}, young massive O and B stars \citep[e.g.][]{daflon04, rolleston20, braganca19}, young late-type stars \citep[e.g.][]{Padgett96, Cunha98, Biazzo11}, Cepheid variable stars \citep[e.g.][]{pedicelli09, lemasle07, lemasle08, lemasle13, genovali14, luck18,  genovali15,  dasilva22, magic22} or  planetary nebulae \citep[e.g.][]{maciel03, perinotto06, henry10, stanghellini10, stanghellini18}, OCs have broadened our understanding of the processes of thin disc formation and evolution.
%allowing the study of the temporal evolution of the thin disc chemical properties.

The best use of OCs as tracers of chemical evolution is possible when spectroscopic observations at high-spectral resolution are available, from which detailed elemental abundances of cluster member stars can be derived.  
In the past years, many spectroscopic observations have been performed with a single slit or fibre, thus observing one star at a time, as in the BOCCE (Bologna Open Cluster Chemical Evolution) Project \citep{bocce, donati12, ahumada13, donati14} and other studies  \citep[e.g.][]{yong05, yong12, casamiquela16, casamiquela17, casamiquela19}.
Subsequently, the employment  of new multi-fibre instruments such as Hydra-WIYN \citep{vonhippel99} and the FLAMES instrument on the ESO VLT \citep{pasquini02} made it possible to study many members of the same cluster simultaneously.
%The WIYN Open cluster project dates back to the late '90s with its first papers, and nowadays it has reached its paper LXXXV \citep[see, e.g.][]{vonhippel99, sara99a, at21}. 
The exploitation of FLAMES at the VLT has favoured the realisation of many large programmes dedicated to the study of open clusters \citep{randich03, randich05, randich06, randich07, sestito03, sestito04, sestito06a, sestito07, sestito08a, sestito08b}. 
The following years have witnessed an enlargement of the sample of OCs with high-resolution spectroscopic observations, providing innovative information on both the inner \citep[e.g.][]{magrini10, reddy16,jacobson16, casamiquela19} and the outer \citep[e.g.][]{carraro07, yong12} disc.
A few years later, the era of large spectroscopic surveys started, also aimed to complement  the subsequent exploitation of the data coming from the {\em Gaia} satellite, which was launched in 2013 \citep{gaiamission, gaiadr1,  gaiadr2, gaiadr3}. 
Among these the {\em Gaia}-ESO survey \citep{gilmore12,gilmore22, randich13,  randich22}, the only one performed on a 8m-class telescope, put a specific focus on the Galactic population of OCs; {\em Gaia}-ESO targeted OCs in a wide range of ages, distances, masses and metallicities, observing large unbiased samples of cluster candidates, with a well defined selection function \citep{bragaglia22, randich22}.

Although focused on the Milky Way field, the other two large surveys at high/medium spectral resolution, GALactic Archeology with HERMES \citep[GALAH;][]{desilva15} and
Apache Point Observatory Galactic Evolution Experiment \citep[APOGEE;][]{majewski17}, also include several clusters in their sample, which allowed them to address the issue of the radial  distribution of chemical abundances \citep[e.g.][]{donor18, donor20, spina21, myers22}. There is also an important contribution from the lower resolution spectroscopic survey, Large sky Area Multi-Object fiber Spectroscopic Telescope \citep[LAMOST;][]{lamost}, whose data, combined with those from the {\em Gaia} mission, helped define the shape of the gradient \citep[e.g.][]{chen09, zhong20, fu22}. 
The contribution of {\em Gaia} with its elemental abundances measured from spectra taken with the onboard RVS spectrograph is adding further constraints and has already delineated the Galactic gradient with its {\sc DR3} \citep{gaia2022}. 
Many more clusters will be included in the coming years thanks to  instruments dedicated to spectroscopic surveys, such as WEAVE \citep[][]{weave} and 4MOST \citep[][]{2019Msngr.175....3D}. 
A review on the state of the art of the radial  [Fe/H] gradient, obtained combining and homogenising the data of open clusters from the three main high-resolution spectroscopic surveys 
({\em Gaia}-ESO, APOGEE and GALAH) is presented in \citet{spina22}.
However, there are numerous questions that are not completely resolved and on which full consensus has not been reached yet, such as, e.g., whether or not there is a time evolution of the gradient traced by clusters in different age bins and  if  the gradient is flattening or steepening with time;  whether clusters are affected by migration as much as isolated stars; and, finally,  if  there is a global relationship between age and metallicity for clusters or whether such a relationship exists only when considering limited intervals in Galactocentric distance. In this paper, we use the sample of OCs observed by the {\em Gaia}-ESO survey (excluding the youngest ones with ages < 100 Myr) to investigate some aspects of the above mentioned open issues, in particular those related to the abundance gradients: namely, the shape of the abundance gradients and their link with the different nucleosynthesis channels;  their temporal change with its implication for Galactic chemical evolution, the role of  migration and possible effects due to spectral analysis;  and the impact of the disc warp, with its variation with time,  on the shape and evolution of the gradients. 
%Our database of clusters' abundances, obtained homogeneously for  high-resolution spectra, covering a wide spectral range, allow us to avoid combining  different literature sources, which might be critical especially in the analysis of elemental radial gradients \citep[see,][for a discussion of this issue]{netopi16, donor18, ray22}. 
%Moreover, in the present work we adopt  the distances and ages homogeneously derived by \citet{CG20} from {\em Gaia} {\sc dr2} data \citep[see][for the impact of the distance scale]{yong12,donor18, donor20}. 
 The structure of the paper is as follows: in Sec.~\ref{section:abundance_and_sample} we describe the abundance analysis and sample selection, whereas in Sec.~\ref{sec:shape} we analyse the shape of the [Fe/H] and elemental abundance gradients. In Sec.~\ref{sec:evolution} we present the time-evolution of the gradients and in Sec.~\ref{sec:discussion} we discuss its possible origins. 
In Sec.~\ref{sec:true}, we select an unbiased sample of stars to trace the time-evolution of the gradient. Finally, in Sec.~\ref{sec:summary} we provide our summary and conclusions. 

\section{Abundance analysis and sample selection}
\label{section:abundance_and_sample}

\subsection{Elemental abundance determination in the {\em Gaia}-ESO UVES spectra}
\label{subsection:abundancedetermination}

In this work, we adopted the final data release  of the {\em Gaia}-ESO Survey, selecting only the stellar parameters and abundances from the high-resolution UVES spectra (resolving power R$=$47,000), covering the spectral range 480.0$-$680.0~nm.
In particular, we considered only the spectra analysed by the Working Group 11 (WG~11), which is in charge of the analysis of UVES spectra of FGK stars in the field and clusters. The structure of the {\em Gaia}-ESO WGs is described in \citet{gilmore22}.
The data reduction and analysis were performed by the {\em Gaia}-ESO consortium, as described in, e.g., \citet{sacco14, smi14, randich22, gilmore22}. 
The main steps for the analysis of the UVES spectra are: {\em i)} data reduction, including radial and rotational velocity determinations carried out
at Istituto Nazionale di Astrofisica (INAF) for UVES using the FLAMES-UVES ESO public pipeline \citep{sacco14}; {\em ii)} spectral analysis with a multi-pipeline strategy, shared, inside the same WG, among different Nodes, and internally homogenised \citep[see][for details]{smi14, randich22}; 
{\em iii)} homogenisation of the WG results on a common scale, and production of the final database  using, as calibrators, benchmark stars, open and  globular clusters  \citep[see][for a description of the strategy and its application]{pancino17, Hourihane22}. 
The recommended parameters and abundances are distributed in the {\sc idr6} catalogue, including those used in the present work: atmospheric stellar parameters T$_\mathrm{eff}$, $\log g$,  [Fe/H], elemental abundances, and radial velocities (RVs). 
%The final number of stars in the {\sc idr6} database with results from WG~11 is 6877, of which 3613 are field stars, 2794 are target stars in the fields of 96 globular and open clusters, and, finally, 470 ae calibrator stars (both benchmark  and  asteroseismic field stars). 
The catalogue is also publicly available in the ESO archive \footnote{\url{https://www.eso.org/qi/catalogQuery/index/393}}. 

\subsection{The elemental abundances}
\label{sec_abu}
In this work, we  considered a total of 25 chemical elements. including iron,  from different nucleosynthesis processes. 
We use abundances: {\it i)} O, Mg, Si, Ca, Ti, considered representative of the class of the $\alpha$ elements, produced by massive stars \citep[e.g.][]{woosley1995}; {\it ii)} two odd-Z elements, namely Na and Al, which show a similar evolution as the $\alpha$ elements \citep[see][for a description of the main astrophysical sites for the production of these elements]{smi16}; {\it iii)} 
several iron-peak elements: Sc, V, Cr, Mn, Co, Fe and Ni; {\it iv)} Cu and Zn, which have a more uncertain origin: most Cu production on Galactic scales might be due to the weak $s$-process acting in massive stars \citep{romano07}, while  large fractions of Zn at low metallicities may come from hypernovae or pair-instability SNe \citep{koba20}, while the situation at high metallicities is less clear; {\it v)} a number of neutron-capture elements, whose availability is unique compared to other large spectroscopic surveys, thanks to the higher resolution and spectral coverage of {\em Gaia}-ESO.
%The origin of the slow ($s$) and rapid ($r$) neutron-capture elements are widely discussed in \citet{magrini18, viscasillas22, vds22}. 
Among the elements whose origin is dominated by the $s$-process we used Y, Zr, Ba, La, and Ce, and among those dominated by the $r$-process we used Mo, Pr, Nd, and Eu. 

\subsection{The solar scale normalisation}
\label{subsection:solarscale}

Most of the stars observed in our sample of OCs are in the giant phase ($\sim$75\%), specifically they are red giant branch or red clump stars; in some of the clusters  {\em Gaia}-ESO observed also dwarf stars in the main sequence (MS) phase
%in particular in Blanco1, M67, NGC2243, NGC2420, NGC2477, NGC2516, NGC3532, NGC6253, NGC6633, NGC6709, and Rup147 
\citep[see][]{randich22, bragaglia22}. 
To produce solar-scaled abundances, we adopted a normalisation procedure based on the abundances of giant and dwarf stars in the open cluster M67, known to have a composition very similar to the Sun \citep[see, e.g.][]{onehag11, liu16}. 

In Table~\ref{solarabundance},  we show the Solar abundances in {\em Gaia}-ESO {\sc idr6}, the Solar abundances of \citet{Grevesse_2007},  which are the reference for the input model atmospheres and the synthetic spectra in {\em Gaia}-ESO, the average abundances of M67, and those for its  giant ($\log g$< 3.5) and dwarf ($\log g\geq$ 3.5) member stars (from the  {\em Gaia}-ESO {\sc idr6}). For the membership analysis we refer to \citet{magrini21}.
The agreement between the Solar and  average M67 {\em Gaia}-ESO abundances and those of \citet{Grevesse_2007} is very good, within 1-$\sigma$ for most elements. A slightly worse agreement, within 2-$\sigma$, is obtained for some elements (O, Al, Ca, Si,  Cu, Ce). 
%The abundances of M67 giant and dwarf stars, taken separately, allow us to estimate the impact that a single set of normalising values to both giant and dwarf stars would have \citep[see also][]{viscasillas22}. 
The differences between abundances in dwarf and giant stars can be attributed both to physical effects, such as stellar diffusion,  and to spectral analysis issues \citep[cf.][]{onehag14,bertellimotta18, souto19}. 
In our analysis, we adopted the two sets of abundances to normalise, respectively, the sample of giant and dwarf stars in OCs. This choice has some impact for elements more influenced by changes when measured in dwarfs or giants (Mg, Si, Ca, Na,  Al), but also for some of the neutron capture elements (e.g., Zr, La, Ce).

\subsection{The open cluster sample}
\label{sec_sample}

Here we considered  the 62 OCs older than 100~Myr as in \citet{viscasillas22}. Abundances in stars belonging to younger clusters might need a specific analysis \citep{baratella20, baratella21, spina21, zhang21}, since they can be affected by magnetic activity and rotation, influencing the determination of the microturbulent parameter, $\xi$, and, in turn, of the whole set of stellar parameters and abundances. In addition, these clusters represent only the latest instants in the global Galactic chemical evolution, characterised by negligible variations with respect to the total timescale. 
For these reasons, we did not include them in this work. 
For the distribution in age and distances of our sample clusters we refer to \citet[see their Fig.~1]{viscasillas22}, 
%(see their Figure~1), 
and for a general description of the open cluster sample in {\em Gaia}-ESO to \citet{randich22, bragaglia22}. 

For each cluster, we performed the membership analysis as in \citet{magrini21} and in \citet{viscasillas22}, namely for 
41 clusters we used the membership probability provided by \citet{jackson22}, while for the remaining clusters we adopt the analysis of \citet{magrini21}. Both membership analyses are based on the 3D kinematics, complementing the RVs from {\em Gaia}-ESO with proper motions and parallaxes from {\em Gaia} {\sc edr3} \citep{gaiadr3}. 
Since for clusters we rely on the average abundances of all members, it is not necessary to introduce a cutoff on the abundances of each individual member based on the signal-to-noise or errors in parameters. We only discarded  stars with high errors in abundance $\geq$ 0.1 dex.
In addition, for each cluster, we used the interquartile range rule to detect potential outliers that fall outside of the overall abundance pattern. 
%This range is defined by (Q1 - 1.5$\times$IQR, Q3 + 1.5$\times$IQR), where IQR is the difference between the 75$^{th}$ and 25$^{th}$ percentiles of the data, being Q1 the lower quartile, Q2 the median, and Q3 the upper quartile  in which the outlier stars stand out from the main distribution for each cluster.

In the Appendix, we provide in Tables~\ref{tab:average_alpha},   \ref{tab:average_odd}, \ref{tab:average_iron}, \ref{tab:average_s},  \ref{tab:average_r} the global metallicity of each cluster from \citet{randich22} %(with membership obtained from \citet{jackson22})
together with the Galactocentric distances, R$_{\rm GC}$,  and the age \citep{CG20}, and the abundance ratios used along the paper. For clusters whose [Fe/H] was not available in \citet{randich22}, we computed the average [Fe/H] using the cluster members adopted in the present work. In Table~\ref{tab:anumber} we also provide the number of member stars used to compute the abundance ratios per cluster and per element. 

\section{The shape of the radial abundance and abundance ratio gradients}
\label{sec:shape}

The shape of the metallicity gradient is an important observational constraint for defining the  timescales of the formation of the Galactic thin disc, the radial variation of the star formation rate and its efficiency, and that of the gas infall. 
Open clusters have been used to track the gradient for many decades \citep[see, e.g.][]{janes79}, and many papers have highlighted the bimodal nature of the gradient, with a break in abundances at around R$_{\rm GC} \sim$10-12 kpc, beyond which the gradient becomes flatter \citep[e.g.][]{bragaglia08, sestito08a, friel10, pancino10, carrera11, yong12, frin13, reddy16, magrini17, casamiquela19, donor20, ZhangLucatello21, netopil22, spina22, myers22}. 
There is a general agreement for the existence of a steeper gradient in the inner disc and an extended plateau in the outer regions, although there is no substantial difference in the goodness of the fits done  with a single slope compared to those with two slopes.

In Table~\ref{Tab:lit_fehgrad}, we present some recent determinations of the slope of the radial  [Fe/H] gradient with OCs.
%There are many papers that used the open cluster population to trace the radial {\bf [Fe/H] gradient}, and in Table~\ref{Tab:lit_fehgrad} we quote just some recent ones, published in the last three years. 
The global slope of the gradient in the literature  varies from $-0.054$ to $-0.077$~dex~kpc$^{-1}$. Samples that contain mainly inner clusters have a steeper slope \citep[see,e.g.][]{carrera19, spina21, myers22, donor20,  ZhangLucatello21}, while samples reaching the Galaxy outskirts have flat global gradients \citep[see, e.g.][]{spina22, netopil22}. 
The gradient sampled by \citet{gaia2022} is an exception since it is limited to the regions between R$_{\rm GC}\sim$ 5 to 12 kpc, but it is quite flat,  $-0.054$~dex~kpc$^{-1}$. 
%In  this case the number of clusters is much higher, but the spectral range and resolution to measure abundances are lower than in the ground-based samples. 
In the last row of the Table, we put as reference the gradient of Cepheids, computed with the high quality sample of \citet{genovali14}.

\begin{table}

\caption{Examples of recent literature slopes of the  [Fe/H] gradient from open cluster samples.  }              % title of Table
\label{}      % is used to refer this table in the text
\tiny{
\centering  
\scalebox{1.0}{
\begin{tabular}{llll}          % centered columns (4 columns)
\hline\hline % inserts double horizontal lines
Reference.  & slope & Range  & n. objs.\\
Reference.  & (dex kpc$^{-1}$) & (kpc) & \\
\hline \hline
This work &  $-0.054\pm0.004$ & 6-21 & 62 \\
&  $-0.081\pm0.008$ & 6-11.2 & 42 \\
&  $-0.044\pm0.014$ & 11.2-21 & 20 \\
\hline \hline
\citet{gaia2022} & $-0.054\pm0.008$ & 5-12 & 503 \\
\citet{netopil22} & $-0.058$ & 6-21 & 136 \\
\citet{spina22}    & $-0.064\pm0.007$ & 5-24 & 175 \\
\citet{zhang21}  & $-0.066\pm0.005$ & 6-15.5 &157\\
\citet{donor20}  & $-0.068\pm0.004$ &6-16$^*$ &  71\\
%\citet{casamiquela19} & $-0.056\pm0.0011$ & 18 \\
\citet{myers22} & $-0.073\pm0.002$ & 6-11.5 & 94 \\ 
\citet{spina21}    & $-0.076\pm0.009$ & 6-16.5 & 134 \\
\citet{carrera19} & $-0.077\pm0.007$ & 6-14.5$^*$ & 90\\

\hline
\hline
\citet{genovali14} & $-0.051\pm0.003$ & 5-15 & 127\\ 
\hline \hline
$^*$ High-quality samples
\end{tabular}}
\label{Tab:lit_fehgrad}
}
\end{table}

\subsection{The shape of the  [Fe/H] gradient}

In the next sections, we discuss our determination of the shape of the  [Fe/H] and elemental abundance gradients. 
We adopted two different approaches to fit the data: a locally weighted linear regression, estimated using a non-parametric regression method that combines multiple regression models in a k-nearest-neighbour-based meta-model, and  a classical linear regression model fit with its confidence interval. 
The former allowed us to follow better the behaviour of the data, while the latter  allowed us to compare with linear fits available in the literature, and to compare with our results. 

\begin{figure}
  \resizebox{\hsize}{!}{\includegraphics{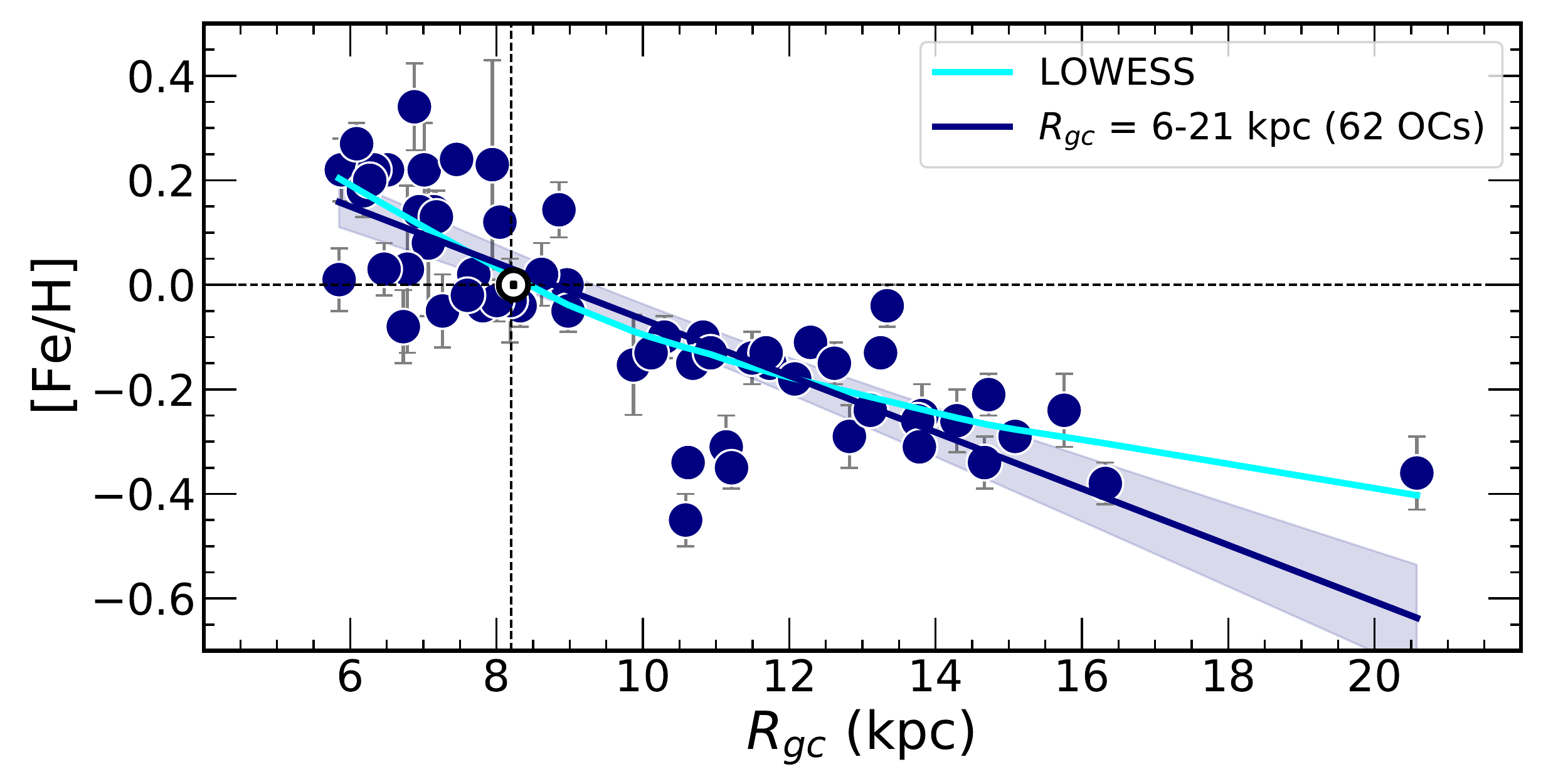}}
  \resizebox{\hsize}{!}{\includegraphics{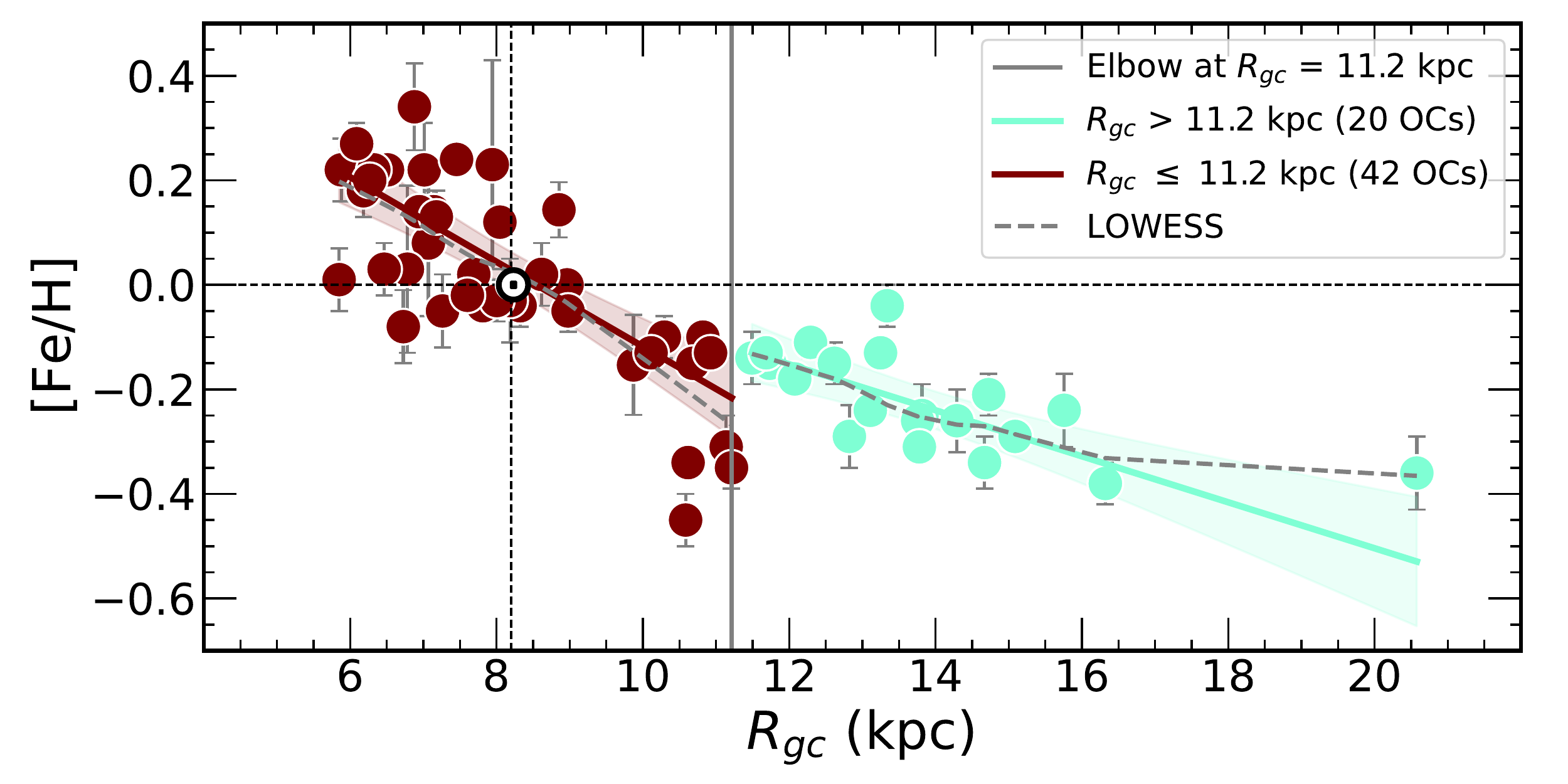}}
  \caption{Radial  [Fe/H] gradient: the global shape. In the upper panel, [Fe/H] versus R$_{\rm GC}$ of our sample clusters;  the blue line is a weighted linear regression model fit to the global trend and the shaded area shows the confidence interval. The cyan curve is a non-parametric lowess model (locally weighted linear regression). In the bottom panel:  [Fe/H] versus R$_{\rm GC}$ with weighted linear regression model fits in two radial ranges, R$_{\rm GC}\leq$11.2~kpc (in brown) and R$_{\rm GC}>$11.2~kpc (in turquoise). The shaded areas show the confidence intervals of the two fits and the dashed gray curves are the lowess models in the two intervals. The vertical grey line marks the slope change point calculated by the Elbow method. The dashed vertical and horizontal lines mark the location and abundance of the Sun. }
  \label{Fig:gradfe}
\end{figure}

In Fig.~\ref{Fig:gradfe}, we show the radial distribution of [Fe/H] as traced by our sample of OCs, with the results of the two fitting procedures. 
In this figure, we plot together  clusters of all ages. 
There is a non-negligible scatter at each R$_{\rm GC}$, related to several aspects that we will expand on in the next sections. %time evolution, radial migration and  the effect of the warp, the latter two particularly important for older clusters. 
In the upper panel, we present a global fit of the sample, in the whole radial range. There is a clear downward trend, with the inner clusters reaching [Fe/H] up to $\sim$0+0.4, and the outer-disc clusters reaching [Fe/H] $\sim-$0.4. 
The scatter increases in the ``transition" region, from the steeper inner gradient to the flatter one, around 10-12~kpc, but it is also quite high in the inner regions.  
Considering a weighted single slope fit, the shape of the gradient is
\begin{equation}
\rm {[Fe/H]}= -0.054(\pm0.004)\times R_{\rm GC}+0.474(\pm 0.045)   
\end{equation}
with a Pearson Correlation Coefficient (PCC) of $-$0.85. 
In the bottom panel of Fig.~\ref{Fig:gradfe}, we consider a two-slope fit, separating the inner   and the outer regions. 
To choose the radius that separates the two regions we applied the Elbow method, to find the "knee" or critical point of slope change where the curve bends from high slope to low slope. Using the kneedle algorithm \citep{Satopaa11}, we found that the cutoff point is at R$_{\rm GC}\sim$11.2 kpc. The two-slope fitting provides a better representation of the behaviour of the outer clusters, although, being dominated by the inner clusters, the PCCs of the one-slope fit and of the inner part of the two-slope fit are very close, as shown below. 
Considering the two radial regions, R$_{\rm GC}\leq$11.2~kpc and R$_{\rm GC}>$11.2~kpc, we obtain a steeper inner gradient: 
\begin{equation}
\rm {[Fe/H]} = -0.081(\pm0.008)\times R_{\rm GC}+0.692(\pm 0.068) \end{equation}
with PCC=$-$0.86 and a much flatter outer plateau 
\begin{equation}
\rm {[Fe/H]} = -0.044(\pm0.014)\times R_{\rm GC}+0.376(\pm0.178) 
\end{equation}
 with  a correlation coefficient PCC=$-$0.61.   
 Our results are also reported in Table~\ref{Tab:lit_fehgrad}  to compare them with literature results. 
 Our global slope is in good agreement with other works covering large radial regions, such as \citet{netopil22, spina22}. %If we consider 
 Our slope for the inner gradient, $-0.081$~dex~kpc$^{-1}$, is in agreement, within the uncertainties, with the gradients mapping the inner regions \citep{carrera19, spina21, myers22, donor20, zhang21}.

\subsection{The shape of the average radial gradients per nucleosynthesis channel}

Here and in next sections, we follow a purely observational approach to describe the elemental abundance and abundance ratio gradients. In comparing the gradient of the various elements, we expect different behaviour because they are sensitive to the timescales of the element production, and to the variation of star formation histories and efficiencies along the disc extent, in particular the inside-out formation of the disc.  
Chemical evolution models can predict these variations accurately, taken into account, in a quantitative way, the different nucleosynthesis. An example is shown in Fig.10 of \citet{minchev14}. in which the combination of the inside-out formation of the disc and different time scales of the production of iron and of the $\alpha$ elements produces the growth of [$\alpha$/Fe] in the outer parts of the disc. We plan a detailed comparison of the data presented in this paper with the predictions of a chemical evolution model (Molero et al. in prep).

\begin{figure}
   \resizebox{\hsize}{!}{\includegraphics{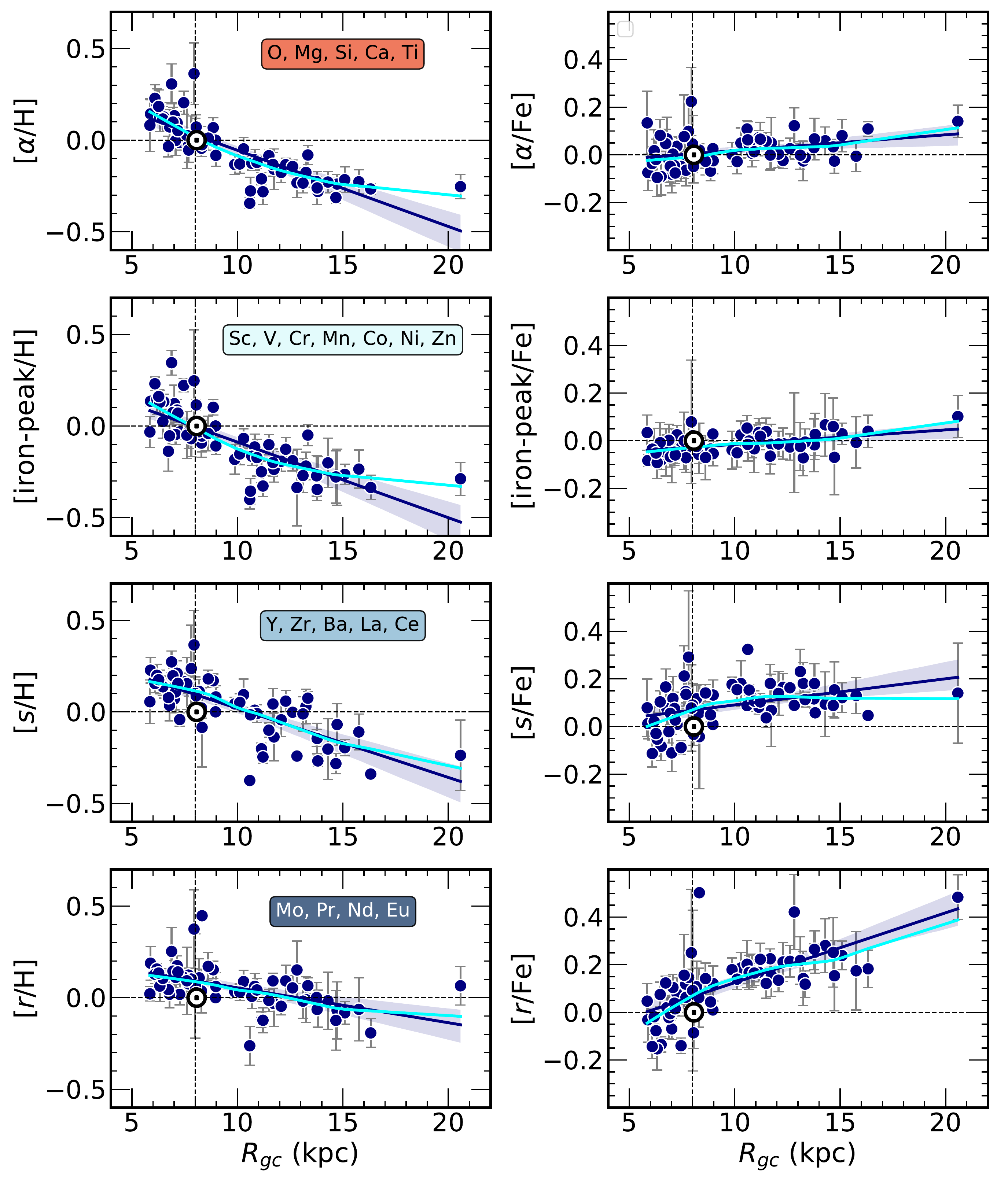}}
  \caption{Radial elemental gradients: average abundances for elements belonging to five nucleosynthesis channels as a function of R$_{\rm GC}$. The blue lines are the linear regression model fits, and the translucent bands around the regression lines are the confidence intervals for the regression estimate. The cyan curves are a non-parametric lowess model (locally weighted linear regression). 
  } 
  \label{Fig:gradaverage_elem}
\end{figure}

In Fig.~\ref{Fig:gradaverage_elem}, we present the radial distribution per group of elements. In each panel, we computed the average abundance of the elements belonging to different groups, characterised by similar nucleosynthesis origins. 
This is clearly a rough approximation because even elements with similar origins, e.g. O and Mg, show differences \citep[see discussion in][]{McWilliam08, vds22}, but this approach can give us an idea of the average behaviour of each element class. 
In the figure, we show  the gradients of both [El/H] and [El/Fe]. 
In the first group, the $\alpha$ elements,  we average the abundances of O, Mg, Ca, Si, Ti.  
In this group,  we include Ti, not properly an $\alpha$ elements, but obtained from $^{48}\mathrm{Cr}$,  its radioactive $\alpha$-element progenitor \citep[e.g., see][]{2003hic..book.....C}.
We exclude 
Na and Al, odd-$Z$ elements synthesised in massive stars as the $\alpha$-elements, whose abundances might be altered  by stellar evolution \citep{smi16}.  
%As an additional caveat, Na and Al are known to
%show star-to-star variations in globular clusters; although no such
%indication has been found in open clusters \citep[][see also \citet{pancino18} for a contrasting view]{bragaglia12, bragaglia14, villanova18}, they might be considered  as
%less secure tracers.
The global $\alpha$-gradient is only slightly flatter than the [Fe/H] one, and, in the [$\alpha$/Fe] versus R$_{\rm GC}$ diagram, it shows a little tendency to rise in the outer Galaxy, combined with a slight descent and high scatter in the inner disc. 
%This behaviour is likely related to differences in the timescales of the production of Fe and of the $\alpha$ elements. Since we considered in the mean value elements as Si, Ca, and Ti, partially produced by SN~Ia, the increase of [$\alpha$/Fe] with R$_{\rm GC}$ is only barely visible. A similar behaviour is observed in the gradient of [$\alpha$/Fe] in the APOGEE sample \citep[][see their Fig. 5]{myers22}. 
In the second group, the iron-peak elements, we considered Sc, V, Cr, Mn, Co, Ni, and Zn.   
For these elements we expect a behaviour similar to the iron one, and indeed the gradient [iron-peak/Fe] is quite flat reflecting the common production sites of Fe and these elements \citep[cf.][]{myers22}. The apparent increase is due only to the outermost cluster. 
In the third group, we combined the five $s$-process elements available in the final data release, namely Y, Zr, Ba, Ce, and La. 
%All these elements are mainly produced by low- and intermediate-mass stars during the asymptotic giant branch (AGB) phase \citep[see, e.g.][]{busso01, cristallo11, karakas16}. 
Their cumulative gradient is slightly flatter than the [Fe/H] one. 
%However, the origin of this flatter gradient is different to that of the $\alpha$-element one. 
%Here, in fact, 
The strong metallicity dependence of the yields of the $s$-process elements comes into play, depressing the production with respect to iron in the inner Galaxy and increasing it in the outer Galaxy. 
%the correct sentence would be that as metallicity increases, the relative production of $s$-elements to iron decreases. This is because at high metallicity the ratio of neutron flux (constant) to Fe seeds (increasing) decreases. 
%There is an intermediate metallicity where the ratio of neutrons to iron atoms is optimal, and then at low metallicity again the lack of Fe atoms decreases the yield  \citep[see, e.g.][]{vescovi21}. 
Finally, in the fourth group, we included the four $r$-process dominated  elements (Eu, Nd, Mo, Pr), which are expected to be produced, in different percentages,  by magnetically driven core collapse SNe or by compact binary mergers, such as neutron star-neutron star, neutron star-black hole and black hole-black hole mergers \citep[see][]{matteucci14}, or by a combination of the two \citep[see, e.g.][]{simonetti19, molero21, vds22}. In both cases, the timescales of their production are short, as indeed shown in their flat [$r$/H] gradient, and in the increasing trend of [$r$/Fe] versus R$_{\rm GC}$. 

\subsection{The shape of the individual element gradients}

\begin{figure}[h!]
  \resizebox{\hsize}{!}{\includegraphics{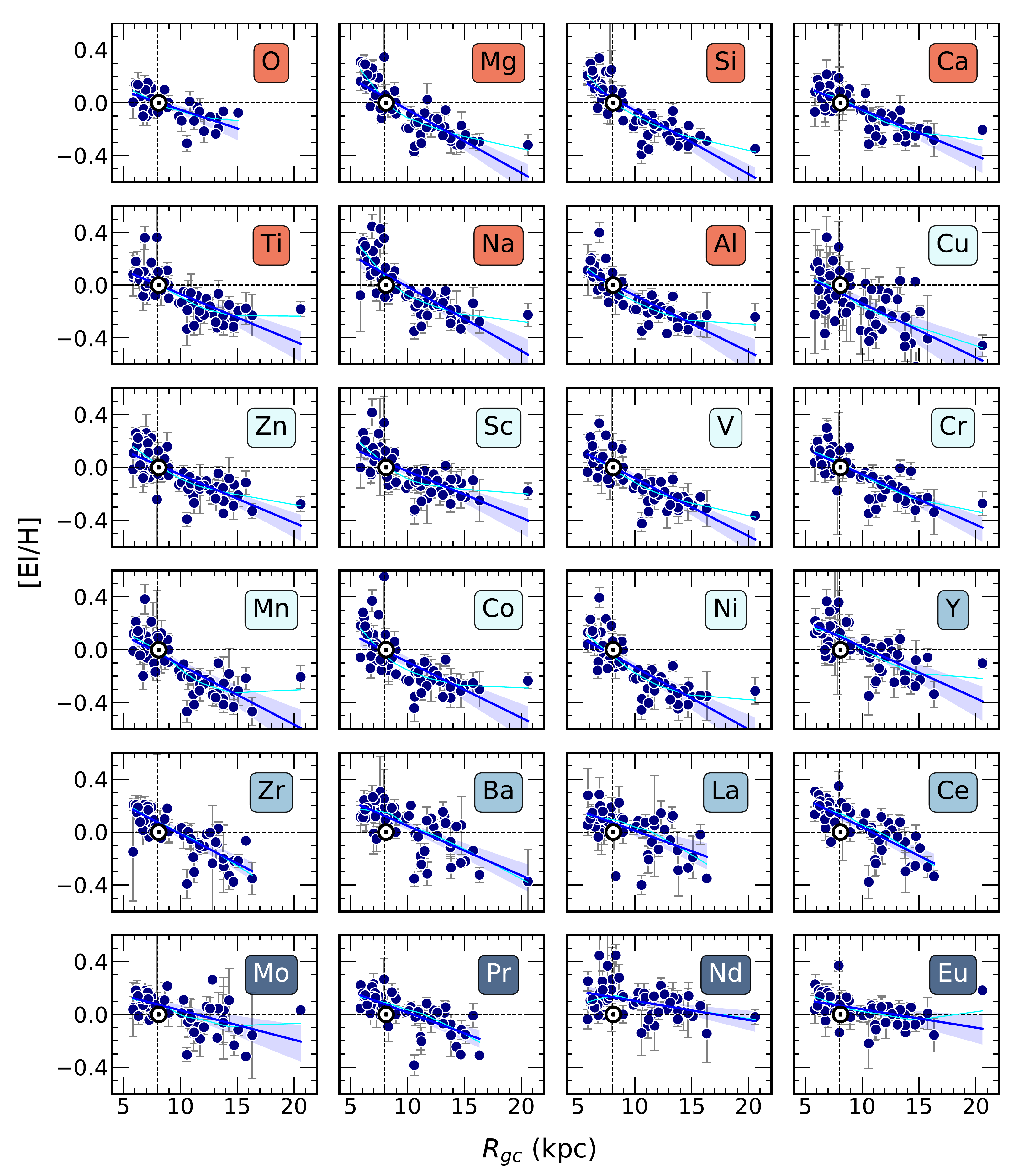}}
  \caption{Radial elemental gradients [El/H] as a function of R$_{\rm GC}$, labelled with different colours  by their nucleosynthesis channel. The blue lines are linear regression model fits, the translucent bands around the regression lines describe a 95\% bootstrap confidence interval, and the cyan curves are non-parametric lowess models (locally weighted linear regression). Each panel is labelled with colours referring to the nucleosynthesis channel (red: $\alpha$ and odd-Z elements; light cyan: iron-peak elements; cyan: $s$-process elements; dark cyan: $r$-process elements.
  } 
  \label{Fig:gradelemenetsh}
\end{figure}

\begin{figure}[h!]
  \resizebox{\hsize}{!}{\includegraphics{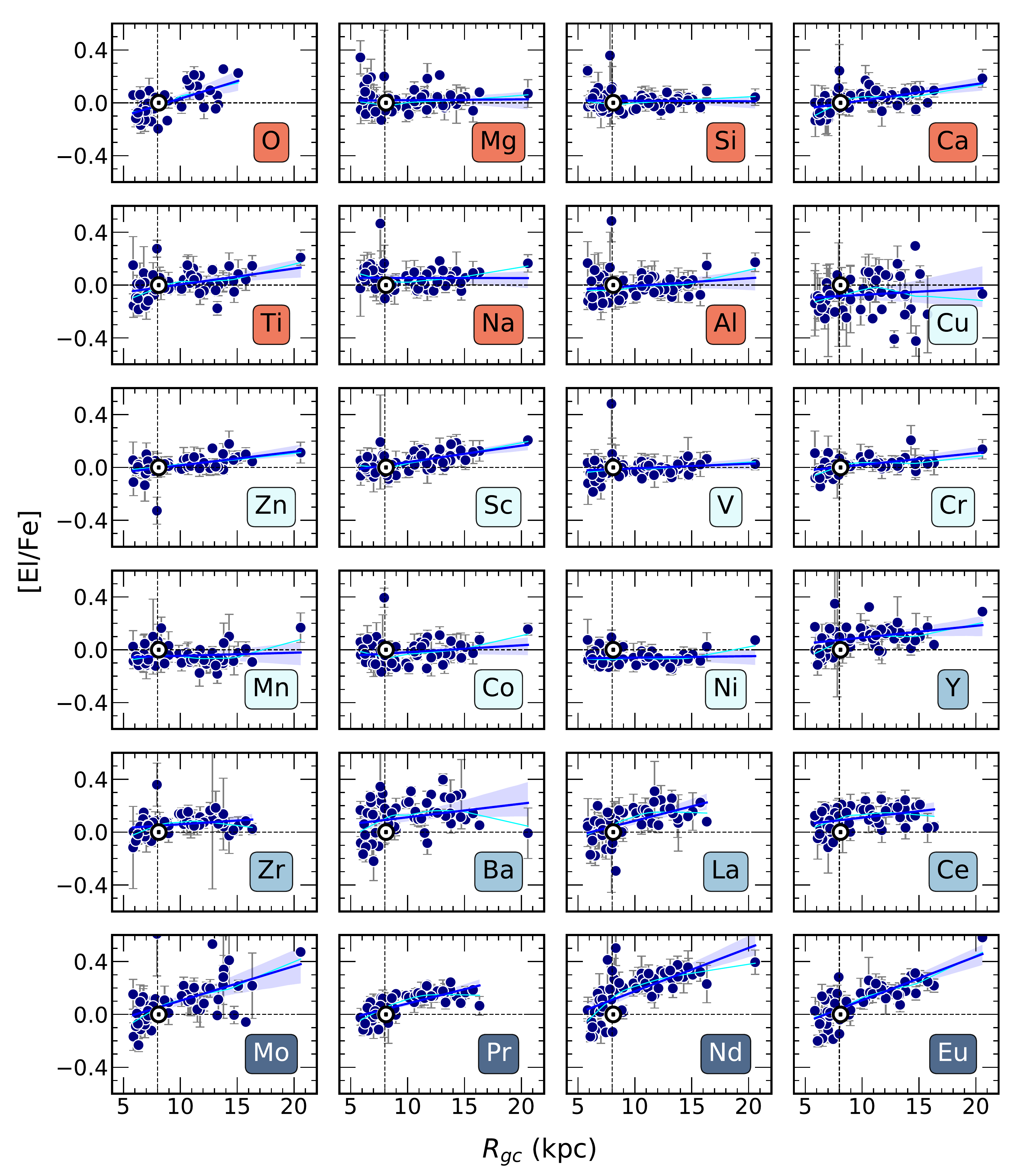}}
  \caption{Radial abundance ratio gradients [El/Fe] as a function of R$_{\rm GC}$.   Symbols and colours as in Fig.~\ref{Fig:gradelemenetsh}.
  } 
  \label{Fig:gradelemenetsfe}
\end{figure}

In Figs.~\ref{Fig:gradelemenetsh} and \ref{Fig:gradelemenetsfe} %(in the Appendix), 
we show the radial [El/H] and [El/Fe] gradients of all the individual elements, grouped by their main nucleosynthesis channels. 
In Tables~\ref{tab:elh_rgc_coefficients} and \ref{tab:elfe_rgc_coefficients} we provide the coefficients of the weighted linear fits and the PCC  of the radial abundance ([El/H]) and abundance ratio  ([El/Fe]) gradients for the whole radial range, and in the two radial regions (R$_{\rm GC}\leq$11.2~kpc and R$_{\rm GC}>$11.2~kpc). Generally, the PCC is much larger for the steeper inner gradients and tends to be lower for the flatter outer gradients.

%\bf There have been many works modelling the chemical evolution of the Galaxy \citep[see][for an excellent review]{matteucci21}. In \citet{romano10} and \citet{prantzos18}, for instance, about the evolution of about twenty elements, from C to Zn, is analysed, finding satisfactory agreement for some elements, and confirming  some well-known problems, such as the underproduction of Mg.
%Among the papers dedicated to the full characterization of the elemental abundance gradient, we recall the paper of \citet{cescutti07} that  modelled the abundance gradients in the disc of the Milky Way for several chemical elements,  adopting an inside-out mechanism  for the formation of the disc, and finding a good agreement with the observations. 

\paragraph{$\alpha$ and odd-Z elements}

In Fig.~\ref{Fig:gradelemenetsh} and Table~\ref{tab:elh_rgc_coefficients},  among the gradients of the first group (O, Mg, Ca, Si, Ti, Na and Al) the gradient of [O/H] (R$_{\rm GC}\leq$11.2~kpc) is the flatter one. 
%There are no clusters in the outer regions with measurements of the oxygen abundances. 
%we need to compare the [O/H] gradient with those of the other elements for R$_{\rm GC}\leq$11.2~kpc. 
In that radial region,  the slope of the [O/H] gradient is also flatter than the one of Mg, the only other element of our sample  expected to be produced only by massive stars.  As discussed in \citet{McWilliam08}, the two elements, although produced both by massive stars,  have different nucleosynthesis mechanisms with a different impact from  stellar winds in their production. Oxygen is produced mostly during hydrostatic burning and expelled through stellar winds. Hence, its yield is expected to have a stronger dependence on the metallicity  than the yields of the other $\alpha$ elements, as Mg,  which are synthesised mostly in the explosive phases. These differences are indeed confirmed by their gradients. 
In addition, for the other elements, e.g. Si, Ca, and Ti, there is a non negligible fraction of their production in SN~Ia  \citep[see, e.g.][]{koba20}. 
The last two elements of the group, Na and Al, are produced by massive stars  as  O and Mg \citep{koba20}, but they have also a non-negligible contribution from low- and intermediate-mass stars, and, as discussed above, might be affected by stellar evolution effects. 
%As shown in \citet{spina21}, there are close similarities between Al and the $\alpha$-elements, observed in the sample of open clusters of GALAH, but also in solar twins \citep{bedell18, casali20}. 
%The slope of the inner gradients from our sample of open clusters, for both Al and Na, are  close to the gradients of [Mg/H] and  [Si/H] and steeper than the [O/H] one. 
In Fig.~\ref{Fig:gradelemenetsfe} we show the behaviour of the radial abundance gradients in the [El/Fe] vs R$_{\rm GC}$ planes.  Among the $\alpha$ elements, [O/Fe]  shows the greatest growth towards the outer regions of the Galaxy,  as expected by the inside-out formation of the Galactic disc. The other $\alpha$  elements and the two odd-Z elements have globally rather flat [El/Fe] gradients,  since they have a non-negligible production in SN~Ia, and thus similar timescales to iron.

\paragraph{Iron-peak elements}
The group of the iron-peak elements (Sc, V, Cr, Mn, Co, Ni, Zn) has a quite homogeneous behaviour in the [El/H]-R$_{\rm GC}$ plane: the slope of their gradients are very similar, around $-0.08$~dex~kpc$^{-1}$  for R$_{\rm GC}\leq$11.2~kpc and
$-0.025$~dex~kpc$^{-1}$   for R$_{\rm GC}>$11.2~kpc (see Table~\ref{tab:elh_rgc_coefficients}). 
In the [El/Fe]-R$_{\rm GC}$ plane, they are almost flat (see slopes in Table~\ref{tab:elfe_rgc_coefficients} and Fig.~\ref{Fig:gradelemenetsfe}). The abundances of Cu~I are the most problematic ones, as shown by the higher scatter of the results. 

\paragraph{$s$-process elements}
The $s$-process elements (Y, Zr, Ba, Ce, and La)  have the most complex behaviour. 
In the [El/H]-R$_{\rm GC}$ plane there is a non negligible scatter in their gradients as traced with clusters of all ages. Their global gradients have typical slopes around $-$0.04~dex~kpc$^{-1}$, close to those of the $\alpha$-element ones. In the [El/Fe]-R$_{\rm GC}$ plane, there is slight increase of the positive slope  in the inner regions (R$_{\rm GC}\leq$11.2~kpc), and an almost flat trend for R$_{\rm GC}>$11.2~kpc, with a decreasing trend  for Zr and  Ba. 

\paragraph{$r$-process elements}
The $r$-process elements (Eu, Nd, Mo, Pr) have mild global gradients in the [El/H]-R$_{\rm GC}$ plane, with typical slopes $\sim -0.02$~dex~kpc$^{-1}$. 
In the [El/Fe]-R$_{\rm GC}$ plane, they have  typical slopes $\sim$+0.03~dex~kpc$^{-1}$, indicating for these elements shorter production timescales than those of iron. In fact, they are mainly produced by magnetically driven core
collapse SNe or by compact binary mergers \citep{matteucci14}, both on shorter timescales with respect then SN~Ia.

\section{Radial abundance and abundance ratio gradients: their time evolution}
\label{sec:evolution}
Several stellar populations have been used to study the time evolution of the gradient: planetary nebulae (PNe), separated in age classes, or combined with abundances of  H~{\sc II} regions \citep[e.g.][]{maciel03, maciel05, stanghellini10, stanghellini18, magrini16}; stars of different ages, in particular those with accurate age determination from asteroseismic observations \citep[e.g.][]{anders17}.
However, compared to other tracers, OCs have the advantage of being a homogeneous population, whose ages and distances can be accurately determined, thereby resulting an excellent tool to follow the time evolution of the abundance gradients.

OCs have been widely used to trace the changes in time of the  Galactic metallicity and of its radial distribution \citep[see e.g.][]{janes79, twarog97, carraro98, friel02, chen03,  magrini09,yong12,frin13, cunha16, spina17,spina21, donor20, zhong20, zhang21, netopil22, myers22}. 
However, a common agreement on the time evolution of the gradient has not been reached so far, either because the cluster samples used in the past were small or because inhomogeneous data were often combined.
The use of homogeneous data from large spectroscopic surveys proved to be fundamental \citep{spina22, myers22}.
In Table~\ref{tab:lit_feh_ev}, we show some recent literature results for the time evolution of the slope of the radial  [Fe/H] gradient from open clusters. 
On average, the gradient shows a limited time evolution, with a slightly flatter gradient for the younger populations. The differences in the results presented in Table~\ref{tab:lit_feh_ev} are mainly due to the differences in the sampled radial regions, in the considered age intervals,  and in which are the younger clusters included in the  samples (see also Table~\ref{Tab:lit_fehgrad}). 

Here, we have the opportunity to revisit this issue with one of the largest samples of open clusters, whose  high-resolution spectra were homogeneously analysed.
%using the same method, including the same set of stellar models and atomic data, and whose age and distances were consistently derived with {\em Gaia} data.   

\begin{table*}
\caption{Literature [Fe/H]  gradients of open cluster samples divided in age bins.  }              % title of Table
\centering  
\begin{tabular}{cccccc}          % centered columns (4 columns)
%\hline\hline % inserts double horizontal lines
Reference & \multicolumn{5}{c}{age bins (Gyr)} \\
%Ref.    &   age bins (Gyr)}    &  &    &     &   \\  
\hline
This work & \multicolumn{2}{c}{0.1<age<1}    & 1$\leq$age$\leq$3 & & age$>$3           \\  
slope (dex kpc$^{-1}$)   &\multicolumn{2}{c}{$-0. 038\pm0.004$} & $-0.063\pm0.006$  & &$-0.084\pm0.019$    \\    % table heading
\hline
\citet{netopil22}    & age<0.4        & 0.4$\leq$age<1  & 1$\leq$age<1.9    &   1.9$\leq$age<4   & 3$\leq$age<5.2  \\  
slope (dex kpc$^{-1}$)      &$-0. 054\pm0.007$ & $-0.058\pm0.008$  & $-0.059\pm0.008$ & $-0.062\pm0.011$ & $-0.081\pm0.031$ \\    % table heading
\hline
\citet{spina21} & \multicolumn{2}{c}{age<1}   & 1$\leq$age <2 & 2$\leq$age<4 & age$\geq$4   \\
slope (dex kpc$^{-1}$)               & \multicolumn{2}{c}{$-0.080\pm0.010$} & $-0.084\pm0.013$     &  $-0.107\pm0.018$ &  $-0.183\pm0.029$ \\
\hline
\citet{gaia2022}  & \multicolumn{2}{c}{age <1}  &1$\leq$age <2 &  2$\leq$age<3 & age$\geq$3  \\
slope (dex kpc$^{-1}$)                 & \multicolumn{2}{c}{$-0.044\pm0.004$} & $-0.065\pm0.010$    & $-0.087\pm0.023$  & $-0.106\pm0.010$  \\
\hline
\citet{myers22}  & age<0.4        & 0.4$\leq$age<0.8  & 0.8$\leq$age<2   &   age$\geq$2  &  \\  
slope (dex kpc$^{-1}$)      &$-0. 052\pm0.003$ & $-0.059\pm0.003$  & $-0.059\pm0.002$ & $-0.052\pm0.002$ &  \\ 
\hline
\hline
\end{tabular}
\label{tab:lit_feh_ev}
\end{table*}

\subsection{The time evolution of the  [Fe/H] gradient}

\begin{figure*}
  \resizebox{\hsize}{!}{\includegraphics{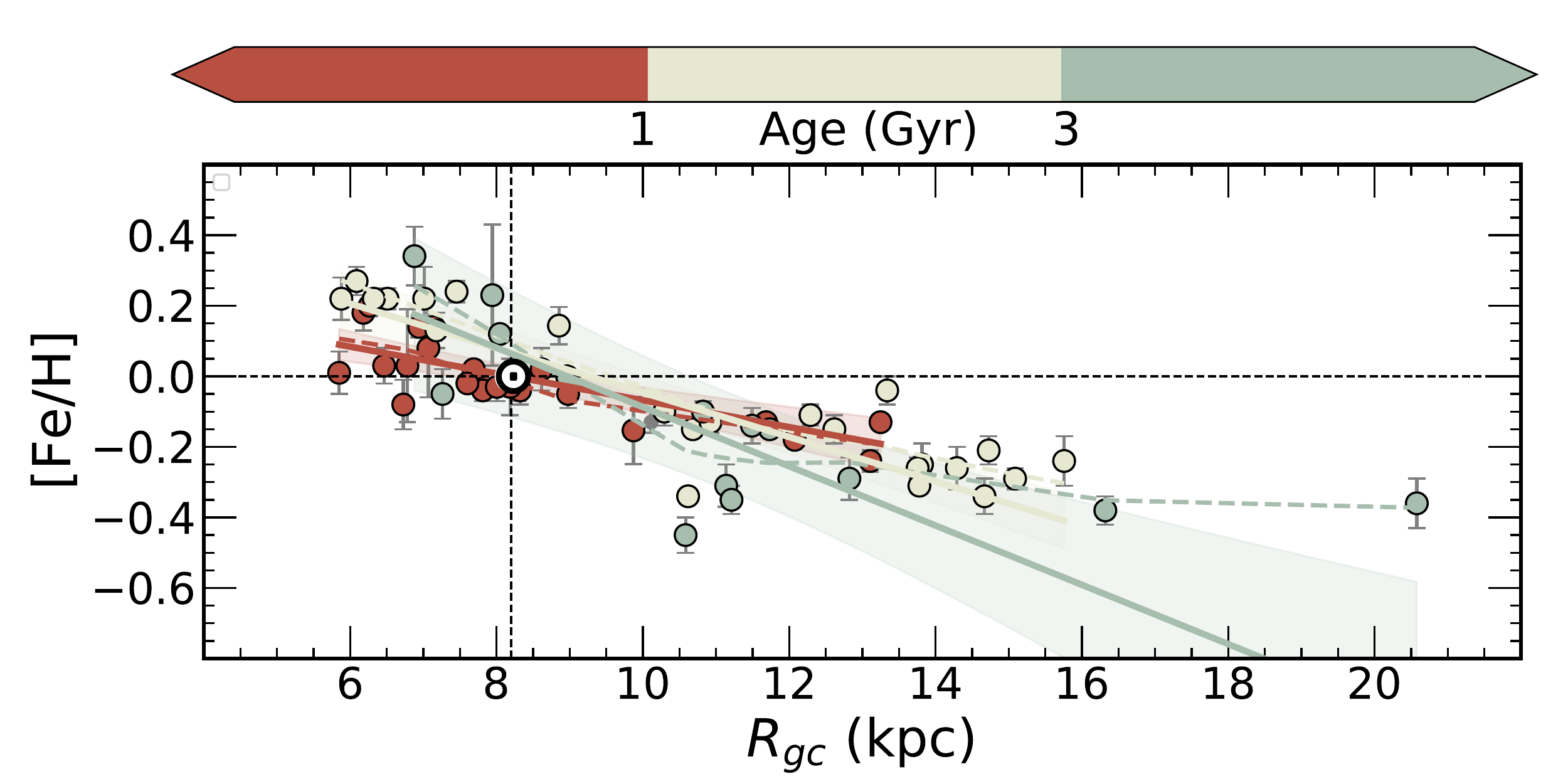}}
  \caption{Radial [Fe/H] gradient: time evolution. The circles are colour-coded by cluster ages. The three solid lines are the weighted linear regressions and the dashed curves are  nonparametric lowess models (locally weighted linear regression) for the three age bins: age< 1~Gyr (red), 1~Gyr$\leq$age<3~Gyr (beige), and age $\geq$3~Gyr (green).
  } 
  \label{Fig:gradfe_age}
\end{figure*}

In Fig.~\ref{Fig:gradfe_age}, we present the radial [Fe/H] gradient of our open cluster sample, divided in three age bins. The age bins were chosen so as to have a similar number of clusters in each one.
In Fig.~\ref{Fig:gradfe}, the dispersion observed in the global  [Fe/H] gradient, in particular in the inner part of the Galaxy, is mostly due to the mixing of clusters of different ages. For instance, between R$_{\rm GC}$ 6.5-7.5 kpc the global  [Fe/H], including clusters of all ages, has a typical dispersion  $\sigma$ = 0.13 dex. 
The dispersion in metallicity of clusters belonging to each one of the three age bins, at a given R$_{\rm GC}$, is smaller than the dispersion of the global sample (see Fig.~\ref{Fig:gradfe_age}). 
From  Fig.~\ref{Fig:gradfe_age} we see a remarkable evolution of the gradient with respect to what  is suggested on the bases of simple chemical evolution arguments, i.e., the oldest populations should be less enriched than the youngest ones.
In the radial range from R$_{\rm GC}\sim$6 to about 10~kpc 
the figure evidences a weird evolution of the gradient, with the youngest clusters more metal poor than the oldest ones. In addition, the gradient  is flatter for the youngest clusters, while  for R$_{\rm GC}>$10~kpc the trend is reversed, and younger clusters have similar [Fe/H] to those of the  older ones. 
Similar conclusions were found in previous works based on open clusters \citep[][but see \citet{netopil22} for a contrasting result]{carraro98, chen03, friel02, magrini09, spina16, randich22}. 
%However, in the compilation of \citet{netopil22}, the youngest clusters (age < 0.4~Gyr) show a gradient with a similar slope as the older clusters. 

%As already mentioned, we only included clusters with age > 0.1 Gyr in the sample, excluding clusters for which standard spectral analysis may provide problematic results \citep[see][]{baratella20, spina20}.
%Hence, from an observational point of view, the results in the inner part of the Galaxy seem apparently robust, and suggest a clear variation in the gradient of the younger populations ($-0.038\pm0.004$~dex~kpc$^{-1}$) compared to those with ages > 1 Gyr, which have a steeper gradient ($-0.063\pm0.006$~dex~kpc$^{-1}$ for 1~Gyr$\leq$age<3~Gyr). 
%A similar, though less pronounced, result was obtained also from the APOGEE  sample \citep{myers22}, in which the younger clusters (age < 0.4~Gyr) have a flat gradient with slope $-0.052$~dex~kpc$^{-1}$ and an intercept around $+0.1$~dex, while the clusters with 0.8~Gyr<age<2~Gyr have a slope $-0.059$~dex~kpc$^{-1}$ and an intercept around $+0.25$~dex.

Also {\em Gaia} {\sc dr3} data  showed a similar  behaviour: \citet{gaia22_parameters} presented two samples, the former composed by young field stars and the latter by stars in young open clusters. In both samples,  they found a flattening of the gradient at recent epochs (a slope of $-0.036\pm0.002$ dex kpc$^{-1}$) in the sample of young and massive stars, to be compared with a slope of $-0.055\pm0.007$~dex~kpc$^{-1}$ in the entire sample, in excellent agreement with our results (see Table~\ref{tab:lit_feh_ev} for more details about the gradient derived only with open clusters). 
Therefore, \citet{gaia22_parameters}  concluded that, from purely {\em Gaia} data, both open clusters and field stars show a flattening of their radial  [Fe/H] gradient in the more recent epochs of Milky Way evolution. 
On the other hand, the works of \citet{spina21, netopil22, myers22} do not indicate large variations with time of the gradient slope (see Table~\ref{tab:lit_feh_ev}), although   the youngest clusters have lower [Fe/H] than the oldest ones. 
We discuss some hypothesis on the origin of the evolution of the gradient in Sect.~\ref{sec:discussion}.

\subsection{The time evolution of the average radial gradients per nucleosynthesis channel}

\begin{figure}[h!]
  \resizebox{\hsize}{!}{\includegraphics{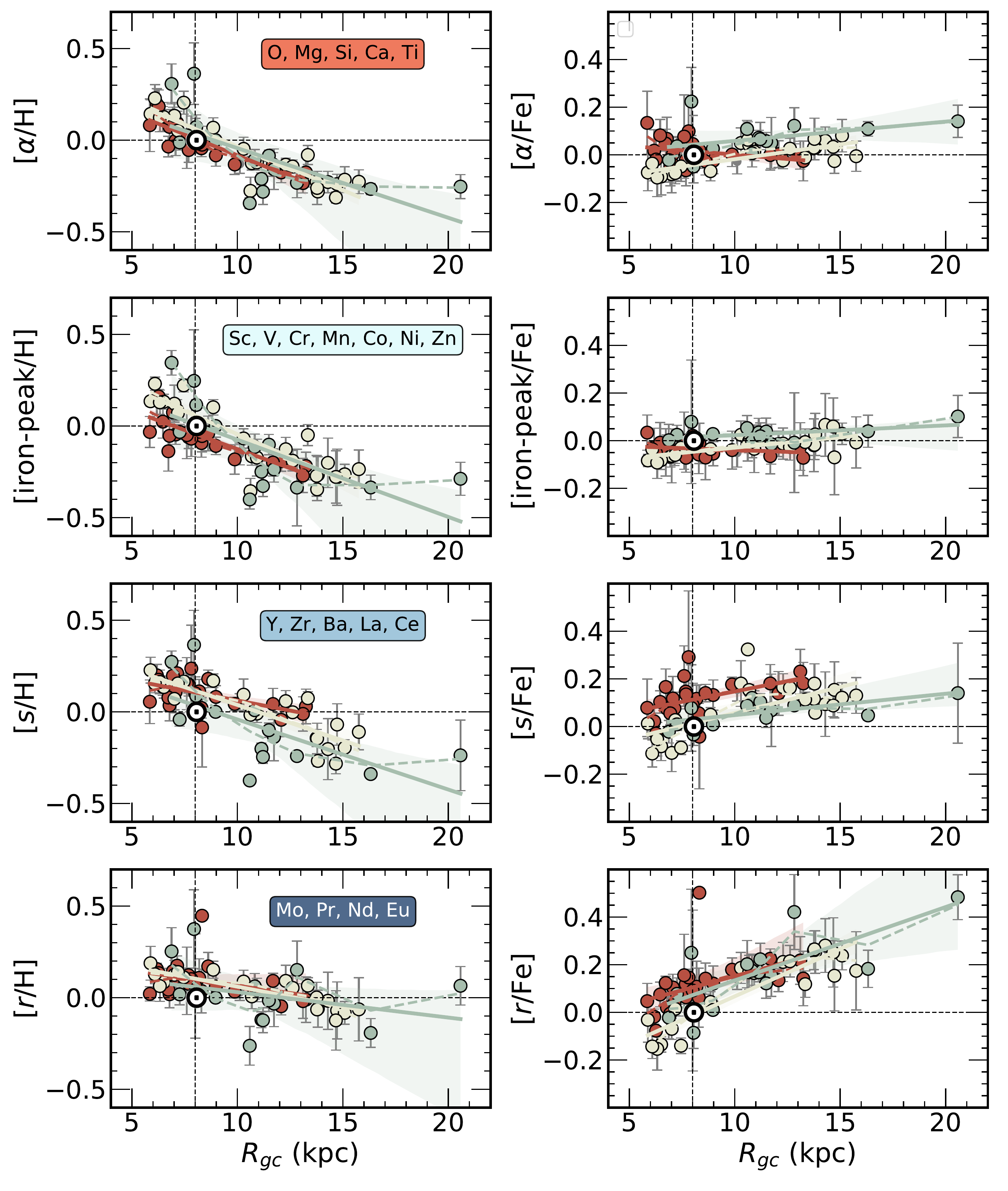}}
  \caption{Radial elemental gradients: average abundances for elements belonging to four nucleosynthesis channels as a function of R$_{\rm GC}$. The lines are the linear regression model fit in every age bin and the translucent bands around the regression lines are the confidence interval for the regression estimate. The dashed curves are non-parametric lowess models (locally weighted linear regression). The symbols and curves are colour-coded by cluster age as in Fig.~\ref{Fig:gradfe_age}.
  } 
  \label{Fig:gradaverage}
\end{figure}

In Fig.~\ref{Fig:gradaverage} we plot the radial abundance and abundance ratio gradients of the averaged elements in the four considered classes divided in the same age bins as Fig~\ref{Fig:gradfe_age}. As above, Na, Al, and Cu are excluded from the average values. 
The [$\alpha$/H] gradient has an evolution with time similar to the [Fe/H] one, but less pronounced, with the oldest clusters only slightly more enriched than the youngest ones. 
In addition, in the [$\alpha$/Fe] versus R$_{\rm GC}$ plot we notice an important difference: for the older clusters (age > 1~Gyr), we have a slight decrease in the inner disc, and a growth in the outer disc, as attested by the inside-out formation of the Galaxy and the different time scales of the formation of the $\alpha$  elements and Fe; for the youngest clusters, 
instead, [$\alpha$/Fe] is flat. %indicating abundances that vary in lockstep.
For the iron-peak elements, we have, as for iron,  a larger enrichment of [El/H] in the older clusters, while the youngest clusters have a flat gradient and are less enriched. 
Due to the common formation sites, [El/Fe] gradients are almost flat at all ages. 
The situation of the $s$-process elements is more complex also with regard to their time evolution. 
The average  [$s$/Fe] gradient  does not show a remarkable variation with age in the inner region, while in the outer regions the youngest clusters are more enriched in $s$-elements. 
%This behaviour is clearly different with respect to the iron one and to those of the $\alpha$- and iron-peak elements. 
This is likely due to the different timescales of their production and to the strong metallicity dependence of their yields \citep{casali20, magrini21b, vescovi21, viscasillas22}.  In the [$s$/Fe] plane the separation between the three age bins is clearer and it is driven by the differences in [Fe/H] of the clusters.
Finally, the $r$-process elements show a behaviour similar to that of the $\alpha$ elements, as expected due to their production on short timescales \citep[see, e.g.][]{matteucci14}.
On the [$r$/Fe] versus R$_{\rm GC}$ plane the separation between the three age bins is likely driven by [Fe/H], as for the $s$-process elements. 

In summary, the average abundance and abundance ratio gradients of the four classes of elements considered have distinct time evolution. 
While for the iron-peak elements we recognise the same behaviour as [Fe/H], $\alpha$ and $r$ process elements have a limited time evolution of their [El/H] gradients, and their slopes remain almost constant with time. The behaviour of the $s$-process elements is the most complex one, and their average gradient might hide this complexity. We describe the gradients element by element in Sect.~\ref{sec:gradelage}.

\subsection{The time evolution of the individual element gradients}
\label{sec:gradelage}

\begin{figure}[h!]
  \resizebox{\hsize}{!}{\includegraphics{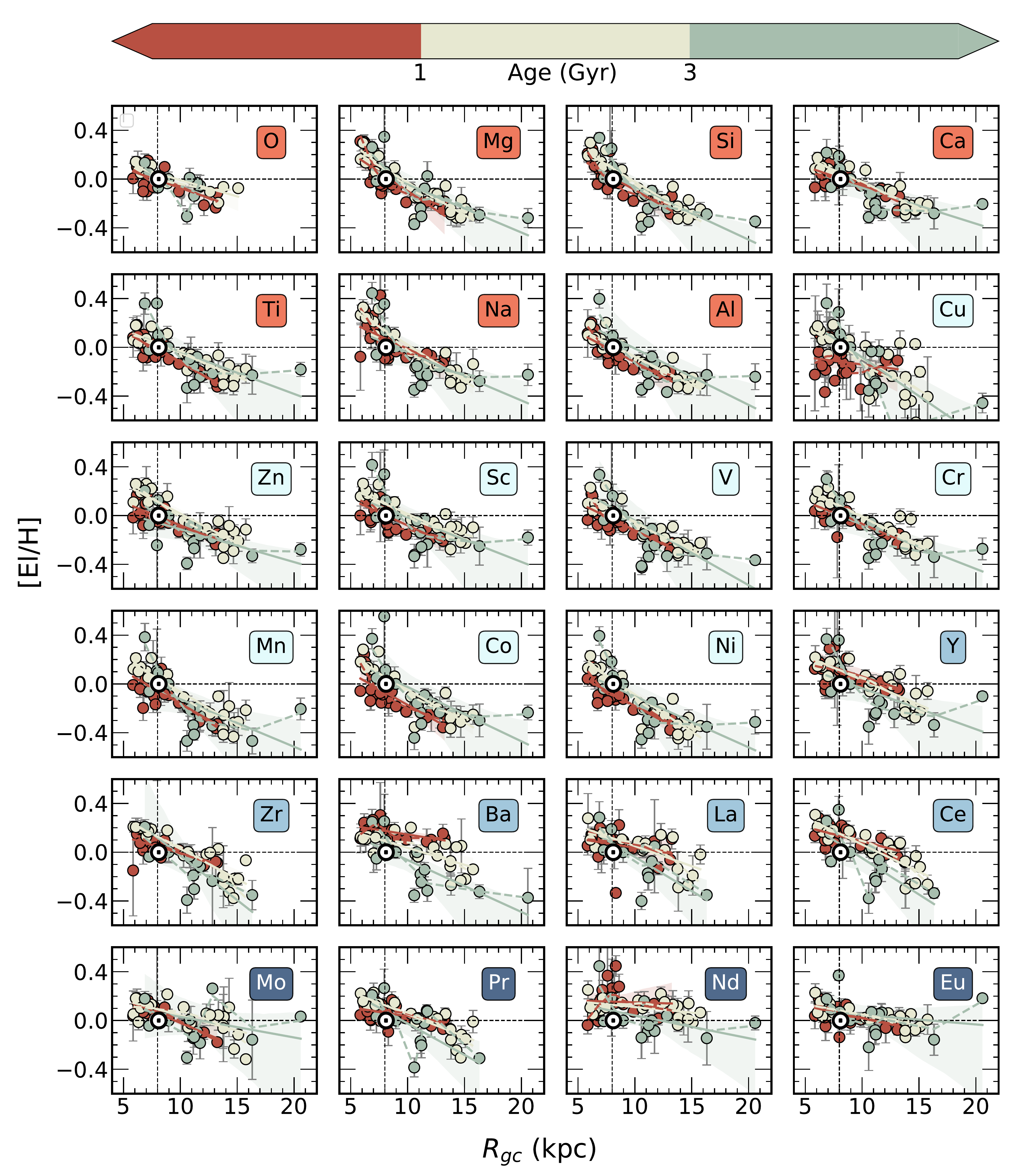}}
  \caption{[El/H] as a function of R$_{\rm GC}$ with clusters divided in three age bins. Symbols and colours are as as in Fig.~\ref{Fig:gradfe_age}.
  } 
  \label{Fig:gradelemenetshage}
\end{figure}
\begin{figure}[h!]
  \resizebox{\hsize}{!}{\includegraphics{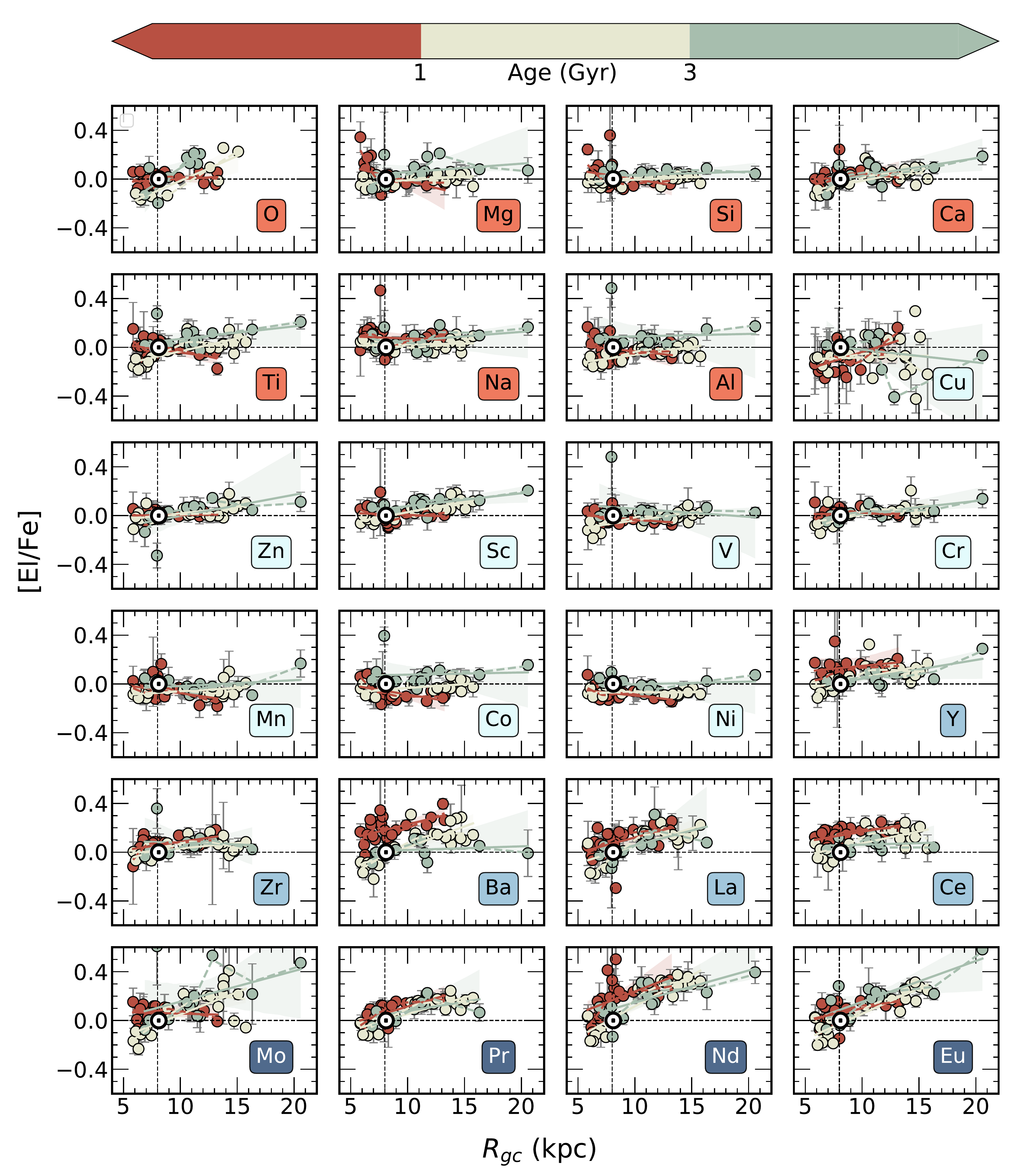}}
  \caption{[El/Fe] as a function of R$_{\rm GC}$ with clusters divided in three age bins.  Symbols and colours are as as in Fig.~\ref{Fig:gradfe_age}.
  } 
  \label{Fig:gradelemenetsfeage}
\end{figure}

\begin{figure}[h!]
  \resizebox{\hsize}{!}{\includegraphics{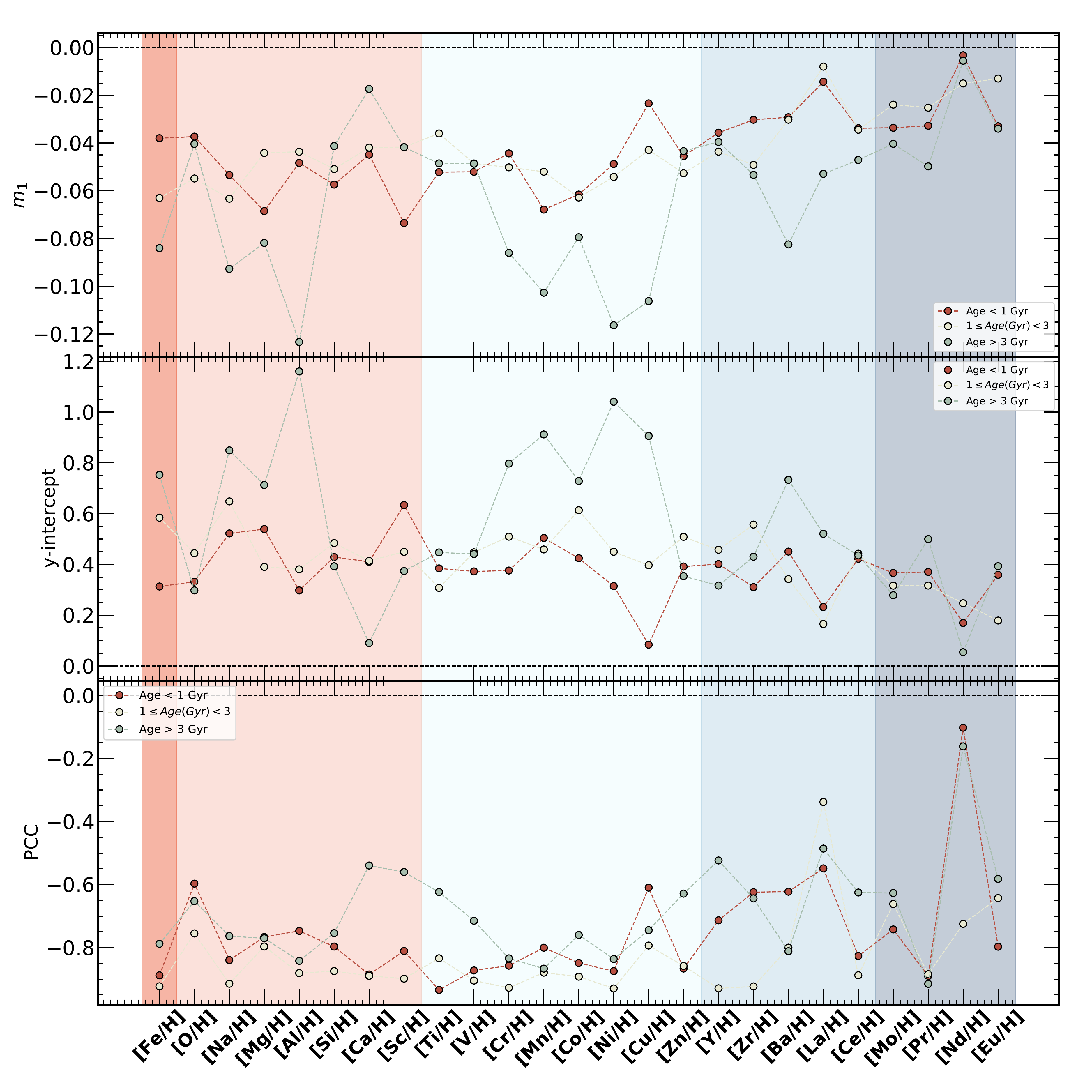}}
  \caption{Coefficients of the weighted linear fits of the abundance gradients (slope- top panel, intercepts -central panel, PCC -bottom panel) } 
  \label{Fig:coefficients}
\end{figure}

In Figs.~\ref{Fig:gradelemenetshage} and \ref{Fig:gradelemenetsfeage}  we present the abundance and abundance ratio gradients of all our sample elements, with open clusters divided in three age bins (0.1~Gyr<age<1~Gyr, 1~Gyr $\leq$ age  $\leq$ 3~Gyr and age>3~Gyr), whereas in Fig.~\ref{Fig:coefficients} we show the coefficients (slope-$m_1$, y-intercept, and PCC) of the linear weighed fits (single slope) in the three age bins for the [El/H] gradients.

%In Figure~\ref{Fig:gradelemenetshage}, we show the time evolution of the abundance gradients of all our sample elements. 

\paragraph{$\alpha$-elements} For the $\alpha$ and odd-Z  elements, our data show a limited time evolution of the slopes and intercepts of their gradients (see Fig.~\ref{Fig:coefficients}). In almost all cases, from O to Al, the shape of the gradient is basically unvaried with time in age range spanned by the youngest bin and the intermediate-age one.  
In none of the $\alpha$ elements  the youngest population appears to be significantly more abundant than the oldest one.  In the two youngest age bins, the values of the intercepts of the gradients are in agreement within the uncertainties.
In Fig.~\ref{Fig:gradelemenetsfeage}, the inner disc clusters at young ages show a flatter trend  for their [$\alpha$/Fe] with respect to the older clusters. The differences are mostly driven by variations in the denominator, Fe. 

\paragraph{Iron-peak elements} In the group of the iron-peak elements, the time evolution is more marked: for Zn, Sc, V, Cr, Mn, Co, and Ni, the youngest clusters appear to be less enriched and they show a  flatter gradient (see top and central panels of Fig.~\ref{Fig:coefficients}). The situation of Cu is less clear due to the high uncertainties on its abundance measurements  and different nucleosynthetic origin. 
The [iron-peak/Fe] abundance ratios show almost flat gradients in the light of the common nucleosynthesis of iron and iron-peak elements. 

\paragraph{$s$-process elements} The $s$-process elements are characterised by an inverse trend with respect to the $\alpha$- and iron-peak elements: the youngest clusters tend to have slightly higher or equal to  [El/H] abundances than the older clusters. 
The $s$-process elements are mainly produced during He-shell burning in low-mass stars (1–3~M$_{\odot}$) \citep[see, e.g.][]{busso01, cristallo11, karakas16}, which enrich the ISM at later times. Thus, the younger stellar populations formed from the ISM enriched by the products of low-mass stars will be also more abundant in $s$ elements (see the values of the intercepts in the central panel of Fig.~\ref{Fig:coefficients}).
In the gradients of [$s$/Fe], we recognise the strong age-dependence of those ratios \citep[see. e.g.][]{dorazi09, maiorca11, maiorca12}: unlike all other gradients, for these abundance ratios we see a clear layering, with the younger clusters having higher [$s$/Fe] and the older ones being less enriched. This is particularly evident for [Y/Fe], [Ba/Fe], [La/Fe] and [Ce/Fe], while [Zr/Fe] has a non-negligible $r$-process production and show little time evolution in its [Zr/Fe] gradient. 

\paragraph{$r$-process elements} Finally, similarly to $\alpha$ elements, $r$-process elements also do not show large variations with age for their gradient slopes and intercepts.
For the $r$-process elements, the patterns are similar to the ones of the $\alpha$ elements, and they are mainly driven by the variations in Fe.

%In Figure~\ref{Fig:gradelemenetsfeage}, we show the time evolution of the abundance ratios [El/Fe] versus R$_{\rm GC}$ for all our sample elements. 

\section{On the origin of the gradient time evolution}
\label{sec:discussion}

From a purely observational point of view, our data confirm an extremely interesting result  that is not immediately apparent. 
Focusing on Fig.~\ref{Fig:gradfe_age} for the  [Fe/H] gradient and on Fig.~\ref{Fig:gradelemenetshage} for the individual abundance gradients, we can see a clear time evolution: older clusters (age>1 Gyr) have steeper gradients and reach higher abundances, while younger clusters (age<1~Gyr) have a flatter gradient, with abundances below or equal to those of the older clusters. 
This evolution is peculiar in the context of chemical evolution, in which one expects the chemical content to grow with time (in the closed-box approximation and in general in large galaxies, as our Milky Way, in which the outflow is not considered dominant), and whose timescales generally do not predict such large changes in just one Gyr, especially during the latest phases of Galactic chemical evolution. 
Only a considerable infall of gas with lower metallicity could be advocated to explain the observations from a chemical evolution point of view \citep[see. e.g., the three infall models described in][]{spitoni22}. 
 
In this section, we investigate three  hypotheses to explain the time evolution of the abundance gradients, evaluating some  aspects of them: 
  
\begin{itemize}
    \item[i)] infall of gas with low metallicity that triggered the latest episode of star formation in the thin disc \citep[][]{ruizlara20} from which the youngest clusters were formed \citep{spitoni19, spitoni22};  
    \item[ii)] migration and selective disruption effects as suggested by, for instance,  \citet{anders17}, \citet{spina21} and recently by \citet{myers22}, in which the oldest clusters are more affected by migration, and are preferentially originated in the inner disc where the global metallicity is higher \citep[see also][for a characterisation of the orbits of clusters as a function of their age]{tarricq22}; 
    \item[iii)] artefacts due to spectral analysis in low-gravity giant stars \citep[see, e.g.][]{casali20b, ZhangLucatello21, spina22, carrera22}. 
\end{itemize}

\subsection{The effect of a recent infall}
The work of \citet{Shanahan19} claimed the discovery of compressed diffuse warm ionised medium in the spiral arm, upstream of the major star formation regions.  \citet{soler20} found that the majority of the filamentary gas structures are aligned with the Galactic plane, with some  significant exceptions of HI filaments mostly perpendicular to the Galactic plane. 
Perpendicular filaments correspond to locations in the disc where there is a significant accumulation of HII regions and supernova remnants.
There is, indeed, observational evidence of the presence of gas infall in the Galactic disc. % both in parallel and in perpendicular directions, with some link with recent star formation episodes. 
The gas from outside can sustain the star formation in the spiral arms, and may have some effect on metallicity, diluting by the same amount the abundances of all elements, and resulting in young stars with lower abundances in all elements.
The chemical characteristics of the youngest and innermost clusters in the sample show a composition in $\alpha$ and $r$-process elements very similar to that of the oldest clusters of the sample and a depletion in the abundances of the iron-peak elements. 
These features can be informative  in understanding the nature of the gas from which these clusters might have been formed. The dilution of the interstellar medium (ISM) with pristine gas causes a depletion of the same amount in all elements, as shown, for instance, in \citet{spitoni22} with their three-infall model. In this model, the thick disc originates from a first infall episode and the youngest thick-disc stars reach solar metallicities. A second infall, then, dilutes the gas, leading to a decrease in [Fe/H] and [El/H] while the [El/Fe] ratios remain substantially unchanged until the star formation activity resumes and new generations of SN~II start polluting the ISM. Finally, a third, recent infall episode is required, which explains the chemical properties of the youngest metal-poor massive stars observed by Gaia \citep{gaia2022}, while also being consistent with the recent, enhanced star formation activity spotted out by \cite{ruizlara20} via an analysis of \emph{Gaia} colour-magnitude diagrams. From the composition of our young clusters, it can be inferred that  the gas that formed them should have low metallicity, i.e. [Fe/H] $< 0$ at the solar radius. 
Furthermore, from Fig.~\ref{Fig:gradelemenetshage} it appears that there is a diversification in the evolution with time of the gradients of the various elements. This seems to indicate that, if a large amount of gas infall is requested, it should also  have a highly peculiar chemical composition.
Full hydrodynamical simulations are required  to investigate this scenario on a quantitative basis, both to identify the source of the gas and to explain its peculiar composition.

\subsection{Migration and selection effects}

Whether or not clusters are affected by radial migration as stars is not completely settled. 
Open clusters are more massive than single stars, and thus the effect of the interactions with perturbing structures, as the spiral arms and the bar, might be different and less pronounced than for single stars \citep[cf.][]{zhang21}.
Chemical evolution models have clearly pointed out that stellar migration has a non-negligible effect on the evolution with time of the gradient \citep[see, e.g.][]{minchev13, minchev14}. 
%Assuming a scenario in which the Galaxy formed inside-out, with a steeper metallicity gradient in the past which tends to flatten with time \citep[see, e.g.][]{minchev18, matteucci21}, we can evaluate the effect of migration by simply observing the gradient of stellar populations of different ages \citep[see, also][]{vickers22}. 
\citet{anders17} attempted to explain the phenomenon of the time evolution of the gradient traced by open clusters by attributing the anomaly to the metallicities of the oldest clusters: their high metallicity for the current location was explained on the basis of their inner birth radius and the preferential migration direction from the inner to the outer disc. 
As discussed in \citet{anders17, spina21, myers22,  gaia2022}, metal-rich
old clusters formed in the inner disc can survive only if they migrate
outward where the Galactic potential is less destructive, while those migrating inward are quickly disrupted. Selection effects and migration might indeed affect the older populations, including clusters with age$\geq$3~Gyr, increasing the scatter in the gradient outlined by them and favouring the survival of metal rich clusters moving from inner to outer disc regions. 
%However, the general effect of radial migration is expected to be a flattening the radial metallicity gradient on a long enough timescale \citep[see, e.g.][]{minchev18, quillen18} and in general a steepening is not expected. 
%\citet{chen20} calculated the migration distances for 146 open clusters, finding that about half of them migrated, but both inward and outward. 
However, while stellar migration might have played a role for the oldest clusters, the intermediate-age clusters show a well defined gradient, with small scatter along the whole disc,  suggesting a very low probability that they have all migrated from the inner disc (see Fig.~\ref{Fig:gradfe_age}).  
Hence, the migration of old clusters can explain the higher abundances and the steeper gradient of the older populations, but it could hardly explain the differences between the gradients of the clusters in the  young and intermediate-age bins, which had less time to migrate \citep[see][and references therein]{frankel18}.

\subsection{Artefacts of spectral analysis in  giants with low surface gravity}

\begin{figure}[h!]
  \resizebox{\hsize}{!}{\includegraphics{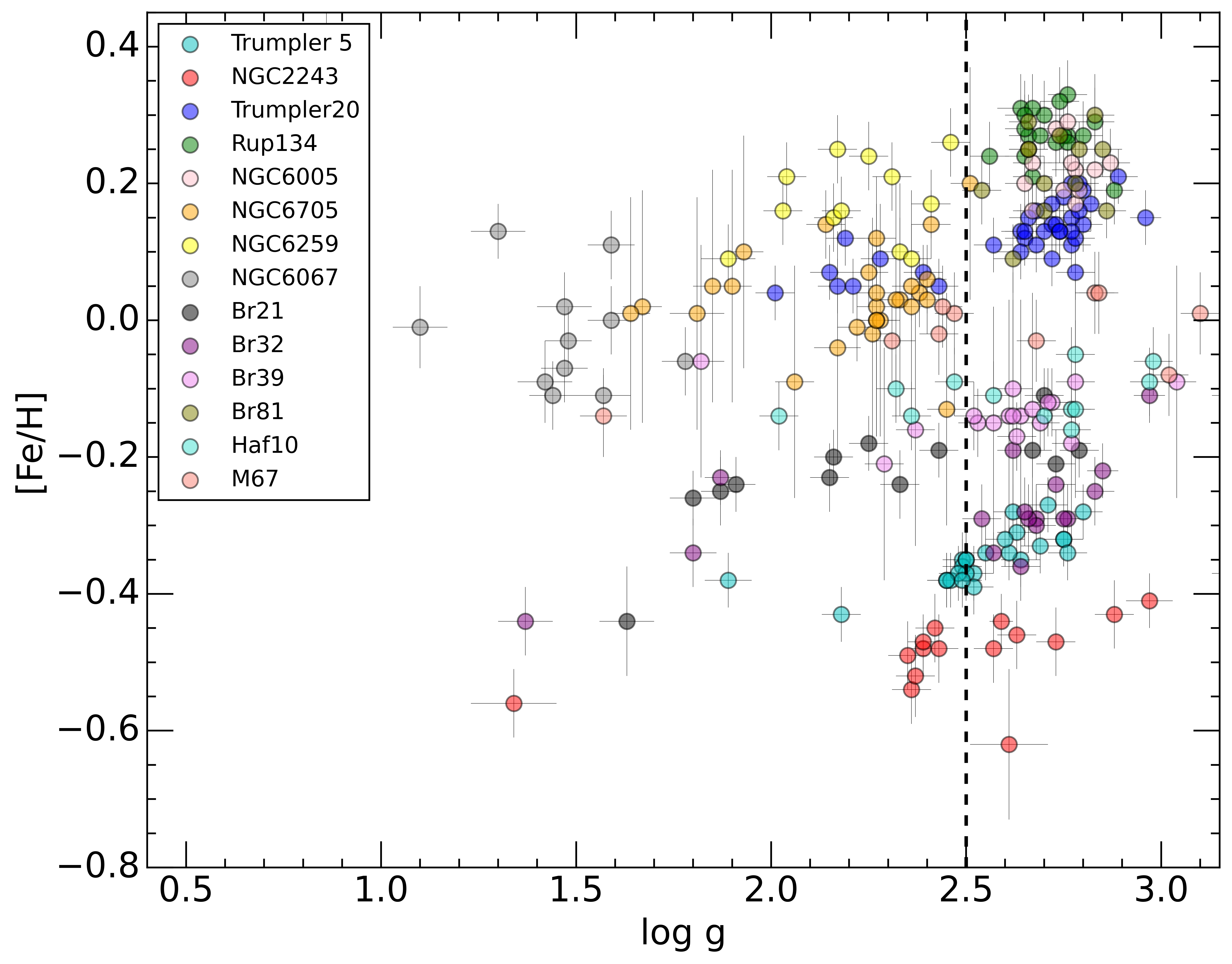}}
  \caption{[Fe/H] as a function of $\log g$ in giant stars members of a sub-sample of OCs. The vertical line marks $\log g$=2.5. 
  } 
  \label{Fig:fehlog}
\end{figure}

The unexpected shape of the gradient in the inner disc outlined by the youngest clusters is driven mainly by the metallicity derived in clusters that host only or mainly giant stars with low $\log g$. 
The giant phase for young massive stars is indeed characterised by high luminosities and low $\log g$. 
In these stars, particularly those with high metallicity, spectral analysis is often prone to strong systematic effects that lead to an underestimation of metallicity the lower the gravity. 
\citet{casali20b} carefully evaluated a number of possibilities to explain the origin of this issue, such as testing the use of photometric parameters instead of spectroscopic ones, the impact of Non Local Thermodynamic Equilibrium (NLTE), the choice of the line list, the use of different atmosphere models, and the definition of the pseudo-continuum.
In Fig.~\ref{Fig:fehlog}, we show [Fe/H] as a function of $\log g$ in a sample of OCs in which we observed  numerous giant stars in a wide range of $\log g$. 
The general trend indicates a decrease in [Fe/H] for stars with lower gravities. A clear example is shown by member stars in the cluster Trumpler~20, where stars with $\log g$>2.5 have [Fe/H] around 0.15, while those with lower gravity have  [Fe/H] between 0 and 0.1. 
Therefore, when the number of stars  with $\log g$>2.5 dominates, we have smaller effects in determining the average cluster abundance. Instead,  in the case of young clusters, in which only  stars with lower gravity are present (see NGC6067, NGC6259, NGC6705), their  [Fe/H]  is most likely underestimated by a factor depending on the typical gravity of the observed stars. The separation shown in Fig.~\ref{Fig:fehlog} at $\log g$=2.5 is quite arbitrary, and it likely varies with metallicity, and it is set just to guide the eye.  In addition, the effect may act differently for specific elements, due to the different photospheric regions in which their spectral lines forms and their relation with gravity.

The trend of [Fe/H] vs $\log g$ is not specific to the analysis performed by {\em Gaia}-ESO but it is also found in the APOGEE and GALAH surveys, as shown in the Appendix (see Figs.~\ref{Fig:galah} and \ref{Fig:apogee}), so it seems to be related to basic aspects, such as the treatment of opacity in model atmospheres or the use of one-dimensional models. 
In Fig.~\ref{Fig:gaia} we also show the comparison with the {\em Gaia} calibrated spectroscopic results which are in excellent agreement with the {\em Gaia}-ESO  data, but which, being calibrated on open clusters, display similar  [Fe/H] for young clusters as the {\em Gaia}-ESO  ones.
In a forthcoming paper, we will analyse both the effects of the one-dimensional and the 3-dimensional approximation, and the possible impact of magnetic activity in younger, lower-gravity giants, by studying also the effects on various elements.

%\begin{figure}
%  \resizebox{\hsize}{!}{\includegraphics{FIGURE/USED/cool_st%ars_xi.png}}
%  \caption{[Fe/H] as a function of $\xi$ in giant stars 5members of a sub-sample of OCs. The vertical line %mark $\xi$=1.8 km s$^{-1}$. 
%  } 
%  \label{Fig:fehxi}
%\end{figure}

\subsection{Giant stars in the youngest clusters}
\begin{figure}[h!]
  \resizebox{\hsize}{!}{\includegraphics{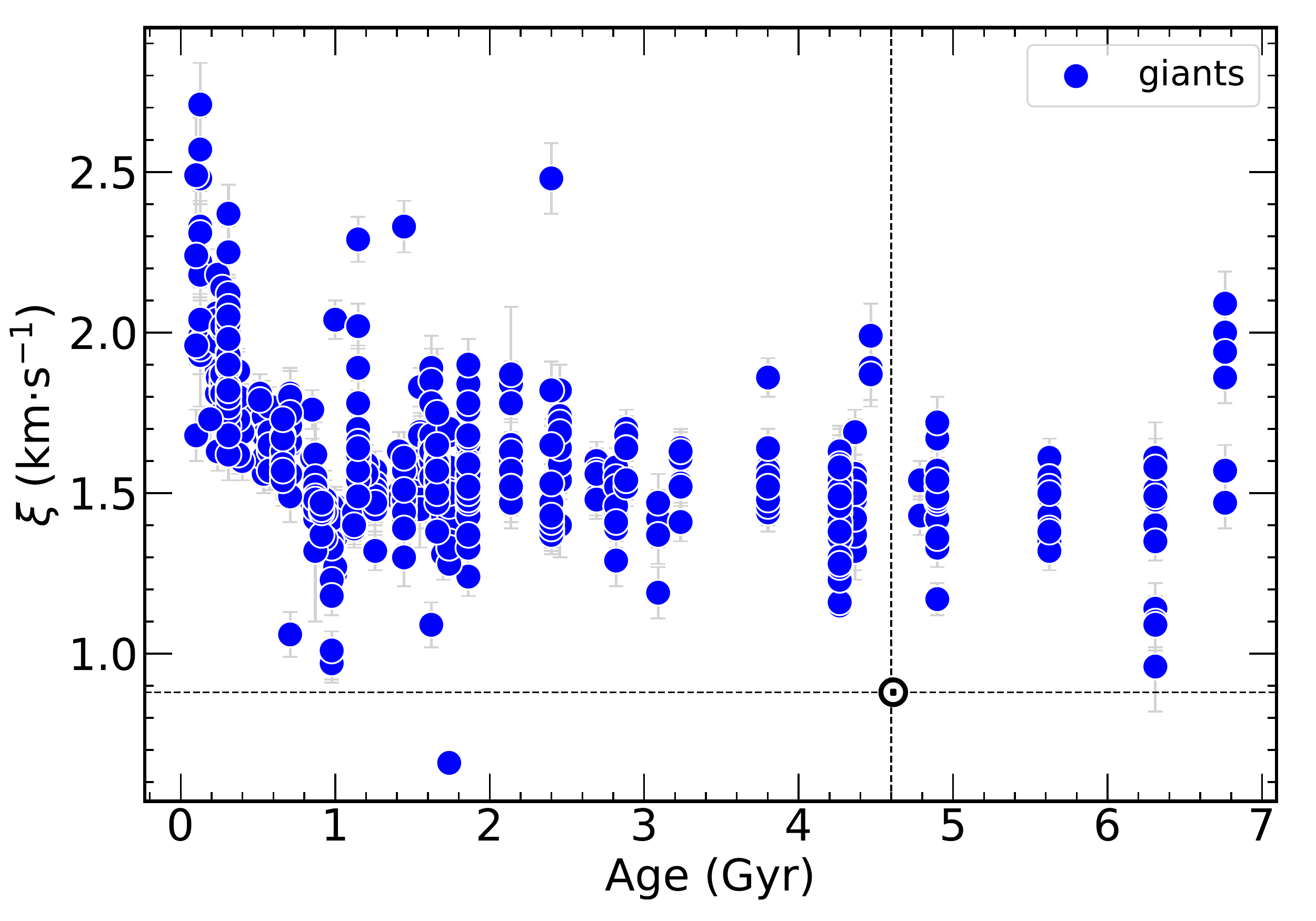}}
  \caption{Microturbulent parameter $\xi$ as a function of the age of the cluster for individual giant cluster members. } 
  \label{Fig:agexigiants}
\end{figure}

In the youngest clusters of our sample,  we also observe  a systematic increase of the measured  microturbulent parameter as shown in Fig.~\ref{Fig:agexigiants}. The origin of this phenomenon is still unknown, however it closely resembles the effect that chromospheric activity has on the analysis of dwarf stars. Recent studies on solar twins have shown that strong absorption lines are intensified at high activity levels \citep{yanagalarza19,spina20}. This effect might be ascribed to the Zeeman broadening of absorption lines that form near the top of the stellar photosphere or even to the presence of cold stellar spots. However, other physical phenomena may be at work \citep{baratella20,baratella21}. In any case, whatever the origin of this effect, the systemic rise of the microturbulent parameter for younger ages may cause biases in the abundance determinations for the youngest stars of our sample. In fact, overestimating $\xi$ implies underestimating both the effective temperature and the abundances of many elements, including Fe. 
To test this effect, we re-analysed the stars of the youngest cluster in our sample, NGC6067. 
A recent work by \citet{alonso17} gives for the giant stars in this cluster an average metallicity  [Fe/H]=+0.19$\pm$0.05 consistent with the Galactic metallicity gradient. In their spectral analysis, the microturbulent parameter  is set at the theoretical value of \citet{adi12}. 
We then re-analysed our sample of stars in NGC6067, keeping $\xi$ fixed at its theoretical value, and using the recommended parameters of T$_{\rm eff}$ and $\log g$ from {\em Gaia}-ESO. In agreement with \citet{alonso17}, we obtain a super-solar metallicity, much higher than the recommended {\em Gaia}-ESO one, in which $\xi$ was derived from the spectral analysis and is higher than the theoretical value.   
Thus, the problem is most likely due to the determination of $\xi$, the physical origin of which might be linked to the magnetic activity of these stars. 

At present, however,  there is scarce information about magnetic activity in giant stars that can lead us to firmly conclude that the effect is related to it.  Recent results seem to support the possibility of a phase of enhanced activity for the most massive giant stars: following \citet{schroder18}, we expect that chromospheric heating undergoes a remarkable reversal and revival as giant luminosity increases, and it becomes more important in young massive giants. In addition, these authors found the effect is enhanced in the most massive (and young) giant stars, which are generally more active. 
Similarly \citet{kostantinova14} found no evidence for magnetic giants below 1.7 M$_{\odot}$, but they detect magnetic G and K giants with M>1.7 M$_{\odot}$ in the upper part of the RGB and in the AGB phase in more than 50\% of their sample. 
In our sample, the youngest clusters with cool giants with $\log g$<2.5 have massed between 3.2 and 4.5~M$_{\odot}$, thus  all possibly hosting magnetic activity.

\section{The `true' gradient}
\label{sec:true}

In this section, we aim at selecting stars for which the standard analysis, as performed by {\em Gaia}-ESO and other spectroscopic surveys, produces  unbiased results. Moreover, we test the role of stellar dynamics, such as the Galactic warp and clusters' orbits. 

\subsection{The effect of biases introduced by the spectral analysis}
\begin{figure}
  %\resizebox{\hsize}{!}{\includegraphics{FIGURE/USED/grad.jpg}}
   %\resizebox{\hsize}{!}{\includegraphics{FIGURE/USED/grad_agege0p4.jpg}}
   \resizebox{\hsize}{!}{\includegraphics{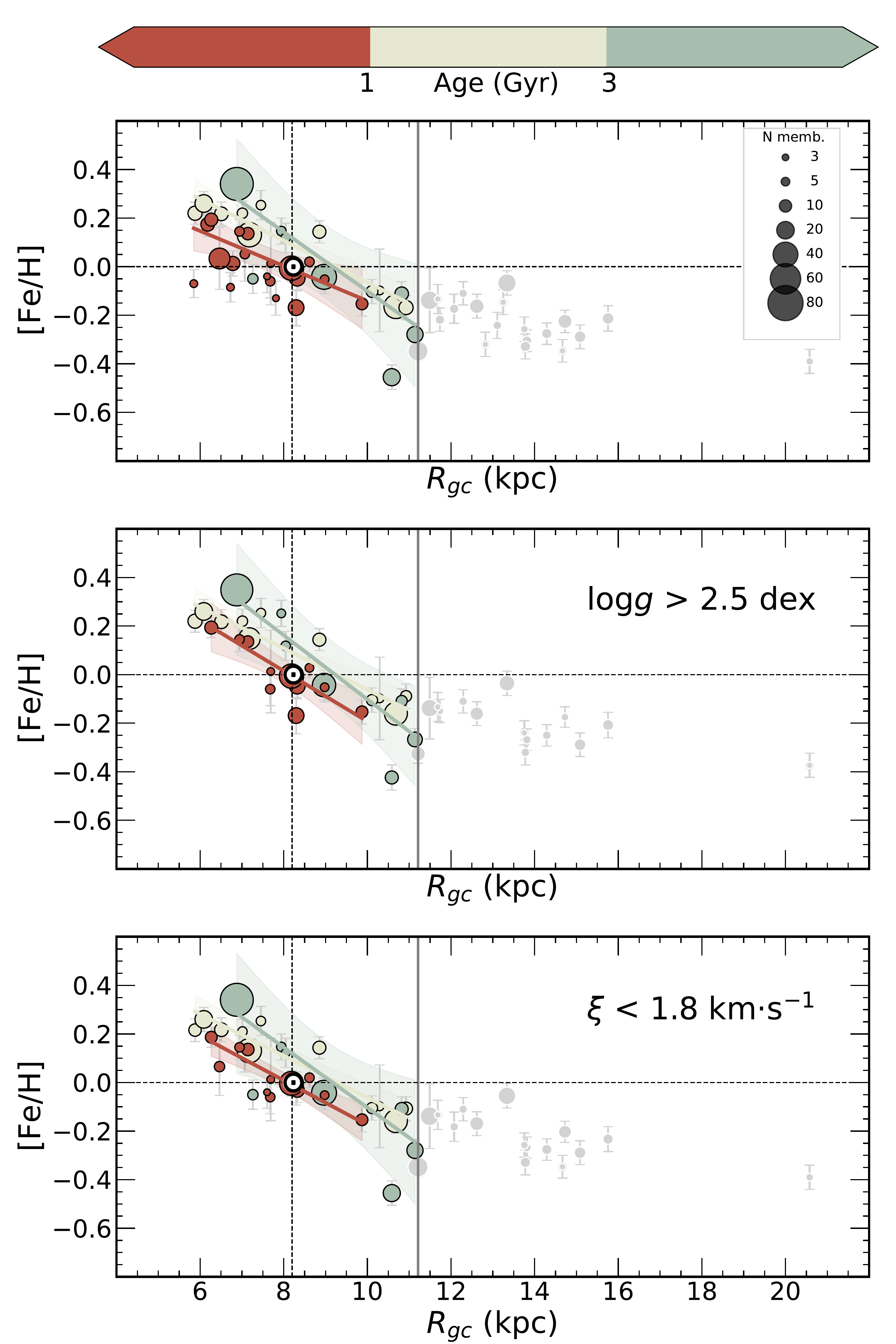}}
  \caption{
  Radial  [Fe/H] gradient with all  sample clusters divided in age bins (top panel), for members with $\log g$>2.5 (central panel) and for cluster members with $\xi$<1.8 km s$^{-1}$ (bottom panel). Open clusters are sized by the number of members and only clusters with 3 or more members are considered. The symbols and curves are colour-coded by cluster age as in Fig.~\ref{Fig:gradfe_age}.} 
  \label{Fig:truegrad}
\end{figure}

In Fig.~\ref{Fig:truegrad}, we show three versions of the gradient, limiting the samples  to the inner radial region: in the top panel, the gradient is drawn considering all clusters (equivalent to Fig.~\ref{Fig:gradfe_age}, but with symbols proportional to the number of members in each cluster and excluding clusters with less than three observed members). In the central panel, we removed giant stars with $\log g$<2.5 in computing the cluster average metallicity. 
In the bottom panel of Fig.~\ref{Fig:truegrad}, we  removed member giant stars with  microturbulent velocity $\xi$>1.8 km s$^{-1}$.  
The results shown in the central and bottom panels are quite similar, since the two removal conditions have similar effect, acting  especially for clusters in which only low $\log g$ stars were observed,  i.e. the youngest ones, which also have high $\xi$. 
The results of the linear fits are also shown in Table~\ref{tab:true}. 
%in which we give the values of the slope, the intercept, and the PCC in the three cases and for the three age bins.
The final effect is a gradient for the youngest populations which is very close to that of OCs with ages between 1 and 3 Gyr, pointing to a very limited evolution of the gradient in this amount of time. The intercept of the gradients of the young clusters rises in the two cases where stars have been removed for their $\log g$ or $\xi$ with respect to the gradient computed with the full sample, thus solving the issue of the lack of enrichment of young populations. The limited evolution with time of the gradient might be a signature of the slow evolution and the lack of major interactions of the Milky Way with other galaxies in the latest few Gyr \citep[see, e.g.][]{sillero17}. The radial  [Fe/H] gradient of their simulated galaxies which are not experiencing strong disturbances evolves, indeed,  smoothly without strong changes. 
The gradient of the oldest clusters (age>3 Gyr) is almost unchanged by the removal of stars with low $\log g$ and high $\xi$. It remains steeper than the one of the youngest populations, and with a positive offset towards higher metallicities, likely because of preferential cluster radial migration from metal-rich regions towards the Galaxy outskirts. 
%So, on the one hand, we have the limits due to the analysis of young stars that prevents the use of stars younger than 0.4 Gyr (but conservatively also < $1 Gyr) analysed in the standard way, on the other hand we have a migration $effect that can become considerable for clusters with ages > 3 Gyr. 

%As stated in \citet{spina22}, the analysis of low gravity giants still remains very elusive and may lead to erroneous conclusions in the study of the gradient. The issue is common to other spectroscopic surveys operating at different spectral resolution and wavelength ranges (see Figs.~\ref{Fig:galah} and \ref{Fig:apogee}).  Conservatively, we recommend to exclude them when calculating the average cluster abundances. 

\begin{table}

\caption{The radial gradients within 11.2~kpc with different restrictions (clusters with more than 3 members). }              % title of Table
\label{}      % is used to refer this table in the text
\centering  
\tiny{
\scalebox{0.95}{
\begin{tabular}{lccc}          % centered columns (4 columns)
\hline\hline % inserts double horizontal lines
All members (38 OCs) & 0.1<age<1    & 1$\leq$age$\leq$3 &  age>3           \\  
slope (dex kpc$^{-1}$)   &$-0. 072\pm$0.016 & $-0.086\pm0.012$  & $-0.122\pm0.025$    \\ 
intercept (dex) & 0.579$\pm$0.121 & 0.795$\pm$0.0998 & 1.119$\pm$0.225 \\% table heading
PCC                      & $-$0.74             & $-$0.82              & $-$0.89 \\
\hline
$\log g$>2.5 (29 OCs)  & 0.1<age<1    & 1$\leq$age$\leq$3 & age>3              \\  
slope (dex kpc$^{-1}$)      &$-0. 104\pm$0.013 & $-0.085\pm0.012$  & $-0.132\pm0.019$ \\
intercept (dex) & 0.847$\pm$0.102 & 0.788$\pm$0.096 & 1.218$\pm$0.167 \\%
PCC                      & $-$0.94             & $-$0.92             & $-$0.95 \\
\hline
$\xi$<1.8 km~s$^{-1}$ (33 OCs) & 0.1<age<1    & 1$\leq$age$\leq$3 &  age>3             \\  
slope (dex kpc$^{-1}$)   &$-0. 092\pm$0.016 & $-0.086\pm0.012$  & $-0.124\pm0.025$ \\
intercept (dex) & 0.745$\pm$0.124 & 0.798$\pm$0.097 & 1.137$\pm$0.229 \\%
PCC                      & $-$0.86             & $-$0.93              & $-$0.89 \\
\hline
\hline
\end{tabular}}
}
\label{tab:true}
\end{table}

\subsection{The effect of the warp}

\begin{figure}[h!]
  %\resizebox{\hsize}{!}{\includegraphics{FIGURE/USED/grad_all.jpg}}
   %\resizebox{\hsize}{!}{\includegraphics{FIGURE/USED/grad_agege0p4.jpg}}
   \resizebox{\hsize}{!}{\includegraphics{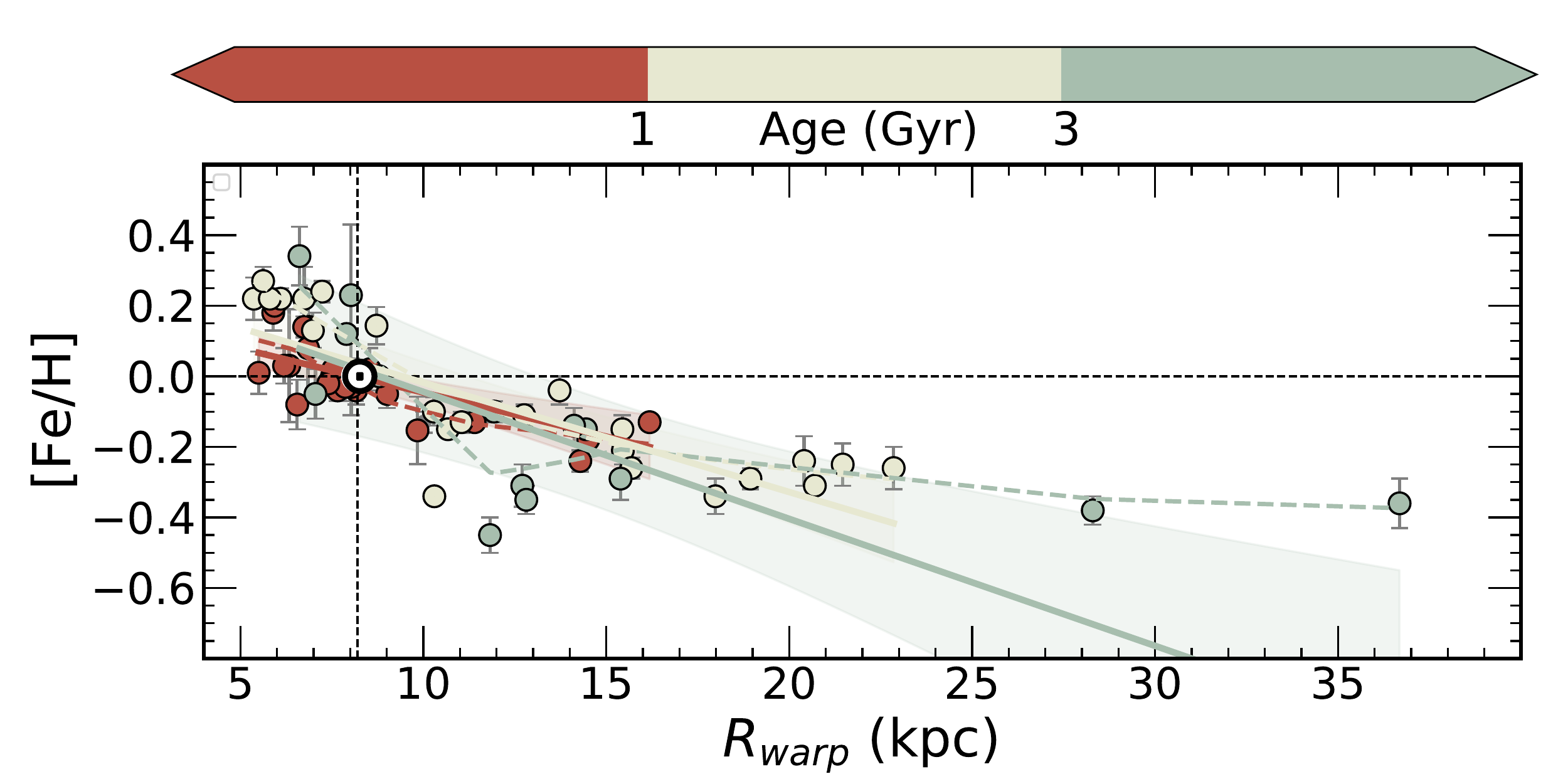}}
  \caption{
  Radial  [Fe/H] gradients with all the sample clusters, colour-coded by the three age bins (one-slope linear weighed regressions, with the confidence levels indicated by shaded areas). Their  R$_{\rm GC}$ are corrected for the time depended warp as in \citet{amore17}.  The dashed lines are the lowess fits. The symbols and curves are colour-coded by cluster age as in Fig.~\ref{Fig:gradfe_age}.} 
  \label{Fig:warpgrad}
\end{figure}

The outer part of the Galactic disk is not flat but it is warping upward in the north and downward in the south. 
%The first detection of the Galactic warp was obtained for the gaseous disc, with observations of H~I \citep{kerr57, oort58} and later in the dust component \citep{Freudenreich94}. Subsequent observations detected the warp also in the stellar component \citep{carneySeitzer93, alard00, drimmel00, drimmel01, lopez02,  lopez20, momany06, reyle09}.
%Kinematic signatures of the Galactic warp are also detected with {\em Gaia} and LAMOST  \citep{Huang18, Poggio18}.
The warp of the Milky Way is not unique, but it is a typical characteristic of the disc galaxies:  about 50\% of spiral galaxies have a warped disc \citep{Reshetnikov98, Guijarro10, Reshetnikov16}. \citet{amores17} studied  the time dependence of the structural parameters of the outer disc, including warp, flare, and disc truncation. They found  evidence of time variation of the thin disc scale length and of the shapes of the warp and flare \citep[see also][]{lopezcorredoira07}. The reasons can be searched in the misalignment between the disc and the  dark halo surrounding the Galaxy or in the interaction with the Magellanic Clouds. As done in \citet{spina22}, we corrected the R$_{\rm GC}$ of our sample clusters considering the effect of the warp and its time dependence. The final effect is that older clusters suffer a larger radial correction than the younger ones, because the disc of the Galaxy was shorter and more curved in the past. In Fig.~\ref{Fig:warpgrad} we show the radial  [Fe/H] gradients in three age bins, with R$_{\rm GC}$ corrected for the time-dependent shape of the warp. 
There is some reduction of the scatter in the region of the change of slope of the gradient, but still some old clusters have a remarkable lower metallicity for their position at R$_{\rm GC}\sim$10-12~kpc. Finally, the time-dependent correction for the Galactic warp has rather unrealistically moved  the two outermost clusters of more than 20 kpc away from the Galactic centre. This is so because the correction for warp of older clusters is based on the model of the shape evolution of the Galaxy of \citet{amore17}, which has considerable uncertainties going back in time.

\subsection{The effect of radial migration (blurring)}
\label{Sec_migration}

We estimated the impact of radial migration, which, among its various effects, can also change the stellar or cluster orbits from circular  to perturbed ones, with a relevant epicyclic component (blurring). 
Stars on perturbed orbits maintain their guiding radius  and their angular momentum,  but they can be found at different R$_{\rm GC}$ along their orbits \citep{bird12}. Local encounters  with molecular clouds \citep[e.g.][]{SS53} or Lindblad resonance scattering between stars and spiral waves  \citep[e.g.][]{SB02} can also produce a change in the angular momentum and consequently in the stellar guiding radius. This effect is more difficult to evaluate, and probably affects to a lesser extent clusters that are more massive than individual stars \citep[cf.][]{Gustafsson16}. 
%{\bf Moreover, radial migration can explain the origin of the most metal-rich field dwarfs ([Fe/H] $>$ +0.25) in the Solar neighbourhood (see Dantas et al., 2022, submitted).}

\begin{figure}[h!]
  %\resizebox{\hsize}{!}{\includegraphics{FIGURE/USED/grad_all.jpg}}
   %\resizebox{\hsize}{!}{\includegraphics{FIGURE/USED/grad_agege0p4.jpg}}
   \resizebox{\hsize}{!}{\includegraphics{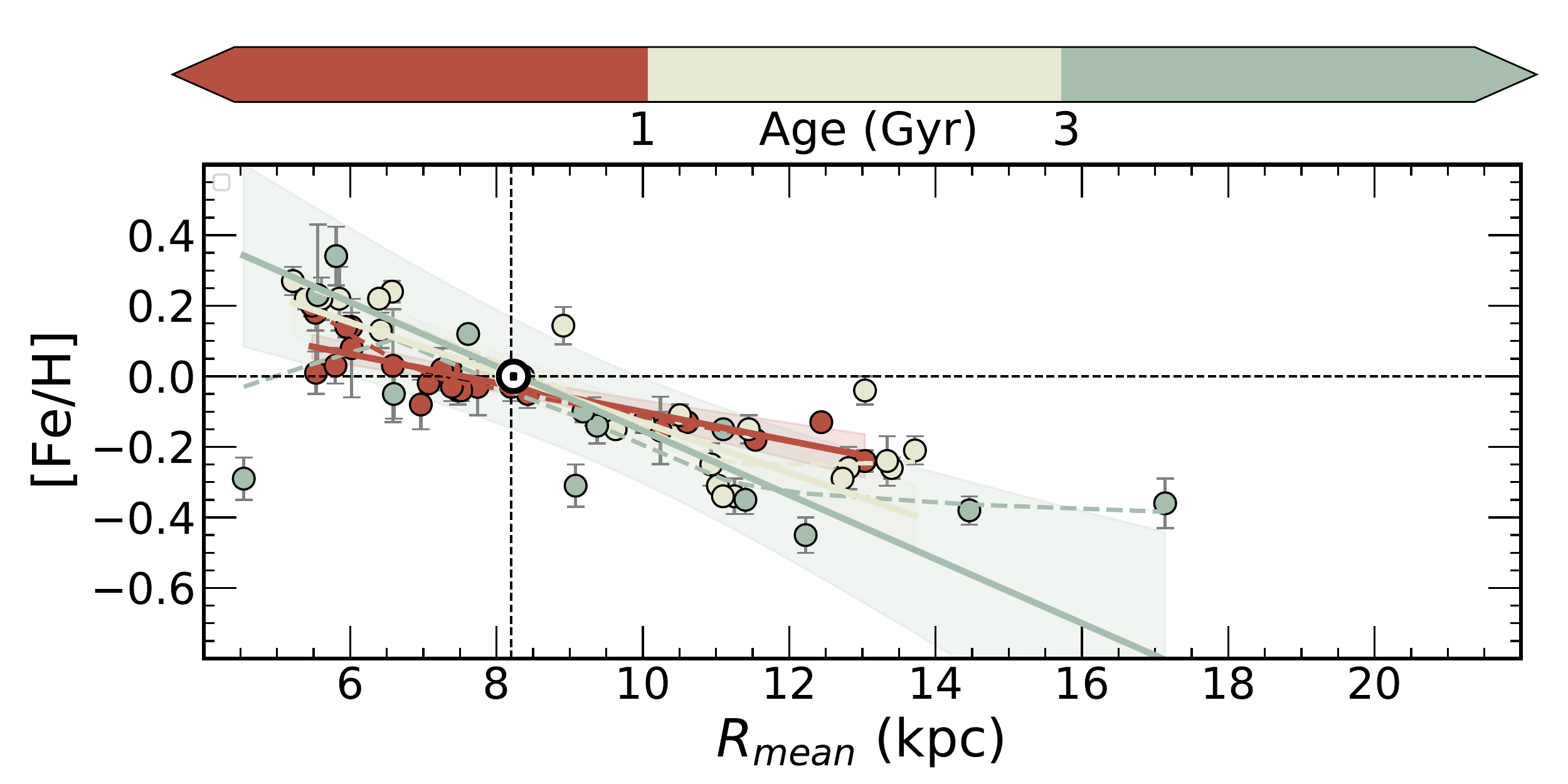}}
  \caption{
  Radial  [Fe/H] gradients with all the sample clusters, colour-coded in three age bins, with  R$_{\rm mean}$ obtained from orbit computation. The lines are the one-slope linear weighed regressions, with the confidence levels indicated by  shaded areas. The dashed lines are the lowess fits. The symbols and curves are colour-coded by cluster age as in Fig.~\ref{Fig:gradfe_age}. } 
  \label{Fig:gradmean}
\end{figure}

We calculated the orbits of our cluster sample using the {\sc galpy} code, with the axis-symmetric potential {\sc MWPotential2014} \citep{bovy15}.
From the clusters' orbits, we computed their guiding radius, R$_{mean}$, defined as the average between the minimum and maximum radius \citep[see, e.g.][]{halle15}. 
We used R$_{mean}$ to recompute the global  [Fe/H] gradients and their time evolution, considering the three age bins. The results are shown in Fig.~\ref{Fig:gradmean}. The high dispersion between 10 and 12~kpc is reduced using R$_{mean}$,  since some clusters located in that regions are likely coming from the outer regions.  

In Table~\ref{tab:grad_corrected_coefficients} we report the three implementations of the  [Fe/H] gradients: with the present-time R$_{\rm GC}$, with R$_{mean}$ and with R$_{warp}$. We provide both the global gradient and the gradients in the three age bins.  
While the correction that takes into account the time-variable warp tends to flatten the gradients, especially for the sample of the oldest clusters, the gradients computed using R$_{mean}$ are slightly steeper. 
%A visualisation of the  impact of the different corrections to the position radius is shown in Figure~\ref{Fig:rmean_rg_rwarp}.
It is likely that the model-dependent correction which should reduce the effect of the warp tends to overestimate the changes in the radius of the outer-disc old clusters, artificially flattening the slope. On the other hand, the gradients computed with R$_{\rm GC}$ and  R$_{mean}$ are consistent within the errors, and also have similar PCC \citep[see][]{myers22}, indicating a secondary role of clusters' migration in shaping the gradient. In the oldest bin, the effect of using R$_{mean}$ improves the PCC of the fit, indicating, as expected,  a more prominent role of migration for the oldest clusters.

%\begin{figure*}
  %\resizebox{\hsize}{!}{\includegraphics{FIGURE/USED/grad_all.jpg}}
   %\resizebox{\hsize}{!}{\includegraphics{FIGURE/USED/grad_agege0p4.jpg}}
%   \resizebox{\hsize}{!}{\includegraphics{FIGURE/USED/Rgc_vs_Rwarp_corrected_v3.pdf}}
%  \caption{
%  Differences among R$_{warp}$, R$_{\rm GC}$ and R$_{\rm mean}$ in our sample clusters. Symbols are colour-coded by R$_{\rm GC}$ as indicated in the legend.   } 
%  \label{Fig:rmean_rg_rwarp}
%\end{figure*}

\begin{table}
\caption{WLS fitting coefficients of [Fe/H] = $m_{1}\cdot$R + c for the global open cluster sample, and  in the three investigated age bins, with R$_{\rm GC}$, R$_{warp}$, and R$_{mean}$. }              % title of Table
\centering  
\begin{tabular}{lrcc}          % centered columns (4 columns)
\hline\hline % inserts double horizontal lines
 & $m_{1}$ & c & PCC \\    % table heading
\hline \hline 
all ages &&&\\
\hline 
R$_{\rm GC}$   & -0.054$\pm$0.004 & 0.474$\pm$0.045 & -0.851 \\
R$_{\rm mean}$ &   -0.060$\pm$0.005    & 0.481$\pm$0.050 & -0.831\\
R$_{\rm warp}$ &  -0.029$\pm$0.003  & 0.255$\pm$0.033& -0.815 \\
\hline\hline 
age (Gyr) < 1 &&&\\
\hline 
R$_{\rm GC}$   & -0.038$\pm$0.004& 0.312$\pm$0.042& -0.888 \\
R$_{\rm mean}$ & -0.041$\pm$0.004& 0.309$\pm$0.035& -0.914 \\
R$_{\rm warp}$ & -0.025$\pm$0.003 & 0.204$\pm$0.035& -0.858 \\
\hline\hline 
1 $\leq$ age (Gyr) $\leq$ 3 &&&\\
\hline 
R$_{\rm GC}$   &-0.063$\pm$0.006& 0.584$\pm$0.064& -0.923 \\
R$_{\rm mean}$ & -0.071$\pm$0.010& 0.579$\pm$0.096& -0.847 \\
R$_{\rm warp}$ & -0.031$\pm$0.004& 0.292$\pm$0.053& -0.871 \\
\hline\hline 
age (Gyr) > 3  &&&\\
\hline 
R$_{\rm GC}$   &-0.084$\pm$0.019& 0.753$\pm$0.186&-0.788\\
R$_{\rm mean}$ & -0.091$\pm$0.015& 0.756$\pm$0.134&-0.872\\
R$_{\rm warp}$ & -0.036$\pm$0.011& 0.316$\pm$0.121&-0.682\\
\hline
\end{tabular}
\label{tab:grad_corrected_coefficients}
\end{table}

\section{Summary and conclusions}
\label{sec:summary}

In this paper, we used the final data release of the {\em Gaia}-ESO survey to investigate the spatial distribution of the abundances and abundance ratios and their time evolution, as traced by open clusters (OCs). 
We can summarise our results in the following points:
\begin{itemize}
    \item[{\em i})] our sample of OCs  trace a well-defined  [Fe/H] gradient, which can be approximated with a single slope fit [Fe/H]=$-$0.054($\pm$0.004)$\times$R$_{\rm GC}$+0.474($\pm$0.045), or with  a two-slope fit, separating the inner region (R$_{\rm GC}\leq$11.2~kpc) and the outer one. Considering the two radial regions, we have an inner gradient, [Fe/H] = $-$0.081($\pm$0.008)$\times$R$_{\rm GC}$+0.692($\pm$0.068), and an outer plateau [Fe/H] = $-$0.044($\pm$0.014)$\times$R$_{\rm GC}$+0.376($\pm$0.178).
    \item[{\em ii})] We derived radial abundance [El/H] and abundance ratio [El/Fe] gradients for other 24 elements. We analysed them both in group considering the average gradients of the $\alpha$-, iron-peak, $s$-process, and $r$-process elements, and per individual element.  All groups of elements and individual elements show negative [El/H] gradients. In the [El/Fe] vs R$_{\rm GC}$ plane, $\alpha$- and $r$-process elements present slightly increasing trends, while iron-peak elements are almost flat. $s$-process elements have a complex behaviour related to their nucleosynthesis. 
    \item[{\em iii)}] We investigated the time evolution of the gradients, dividing the clusters in three age bins: age<0.1~Gyr, 1~Gyr$\leq$age$\leq$3~Gyr, and age>3~Gyr. We found that the [Fe/H] of the youngest bin is characterised by a flatter slope and   for R$_{\rm GC}\leq$11.2~kpc, by a lower metallicity, at a given R$_{\rm GC}$.   For the other elements, iron-peak elements have flatter gradient for the youngest clusters, while for $\alpha$- and $r$-process elements the gradients are almost constant with age. The youngest clusters are more enhanced in $s$-process elements than the older ones, with flatter gradients.   We proposed some ideas to explain this kind of evolution: chemical evolution and infall of metal poor gas; stellar migration and selective disruptive events; finally, we considered the possible problem of the standard spectral analysis of low gravity giant stars, as anticipated by \citet{casali20b, spina22}. We found that in the analysis of  giants ($\log g$ < 2.5) we are prone to underestimate [Fe/H]. In addition, in the youngest giants at low $\log g$, we might also overestimate the microturbulent velocity due to the possibly enhanced magnetic activity.   
    \item[{\em iv)}] We evaluated the effect of the Galactic warp, considering also its shape in the past following \citet{amore17}. The gradients computed with distances deprojected in the warped disc tend to be flatter, especially for the older clusters. We also took into account  stellar migration, computing the orbits of OCs. Using their R$_{mean}$ instead of R$_{\rm GC}$, we computed the gradients. The scatter in the critical region around 10-12~kpc is reduced, but the global shape of the gradient is almost unvaried within the errors. 
\end{itemize}
We conclude that the radial  [Fe/H] gradient as traced by OCs contains much important information on Galactic and stellar physics. Dividing clusters in age bins and carefully  excluding young cool low-gravity giants allow us to trace the time-evolution of the radial metallicity gradient, which is very limited in time indicating a slow and stationary formation and evolution of the thin disc. The older clusters (age>3~Gyr) still show a slightly steeper gradient, with an offset towards higher [Fe/H], likely related to preferential clusters' migration from the inner to the outer disc. 
Upcoming instruments dedicated to spectroscopic surveys, such as WEAVE and 4MOST, will increase the number of clusters with a metallicity determination, but for the moment the {\em Gaia}-ESO data will remain unbeaten in terms of spectral range and resolution, and consequently the number of measured chemical elements.

\begin{acknowledgements}
We would like to thank the referee for her/his careful reading and useful suggestions. 
Based on data products from observations made with ESO
Telescopes at the La Silla Paranal Observatory under programmes 188.B-3002,
193.B-0936, and 197.B-1074. 
These data products have been processed by
the Cambridge Astronomy Survey Unit (CASU) at the Institute of Astronomy,
University of Cambridge, and by the FLAMES/UVES reduction team at
INAF/Osservatorio Astrofisico di Arcetri. These data have been obtained from
the Gaia-ESO Survey Data Archive, prepared and hosted by the Wide Field Astronomy
Unit, Institute for Astronomy, University of Edinburgh, which is funded
by the UK Science and Technology Facilities Council. This work was partly
supported by the European Union FP7 programme through ERC grant number
320360 and by the Leverhulme Trust through grant RPG-2012-541.We acknowledge
the support from INAF and Ministero dell’Istruzione, dell’Università e
della Ricerca (MIUR) in the form of the grant "Premiale VLT 2012". The results
presented here benefit from discussions held during the Gaia-ESO workshops
and conferences supported by the ESF (European Science Foundation) through
the GREAT Research Network Programme. LM, SR, EF, GC, MVdS, GGS, AB acknowledges partial support from Premiale 2016 MITiC.
MLLD and RS acknowledge support by the National Science Centre, Poland, project 2019/34/E/ST9/00133. MB is supported through the Lise Meitner grant from the Max Planck Society. We acknowledge support by the Collaborative Research centre SFB 881 (projects A5, A10), Heidelberg University, of the Deutsche Forschungsgemeinschaft (DFG, German Research Foundation).  This project has received funding from the European Research Council (ERC) under the European Union’s Horizon 2020 research and innovation programme (Grant agreement No. 949173). TB was supported by grant No. 2018-04857 from the Swedish Research Council.
UH acknowledges support from the Swedish National Space Agency (SNSA/Rymdstyrelsen).
GC acknowledges support from the European Research Council Consolidator Grant funding scheme (project ASTEROCHRONOMETRY, G.A. n. 772293, http://www.asterochronometry.eu).
CVV acknowledges support from ''Direzione Generale per la diplomazia pubblica e culturale" of the ''Ministero degli Affari Esteri e della Cooperazione Internazionale" and the  "Istituto Italiano di Cultura in Vilnius".
\end{acknowledgements}

\bibliographystyle{aa}
\bibliography{Bibliography}

%%%%%%%%%%%%% APPENDIX %%%%%%%%%%%%%
\begin{appendix}
\section{Appendix}
\subsection{Effect of the analysis of low gravity giants in other spectroscopic surveys} 

\begin{figure}
\includegraphics[scale=0.3]{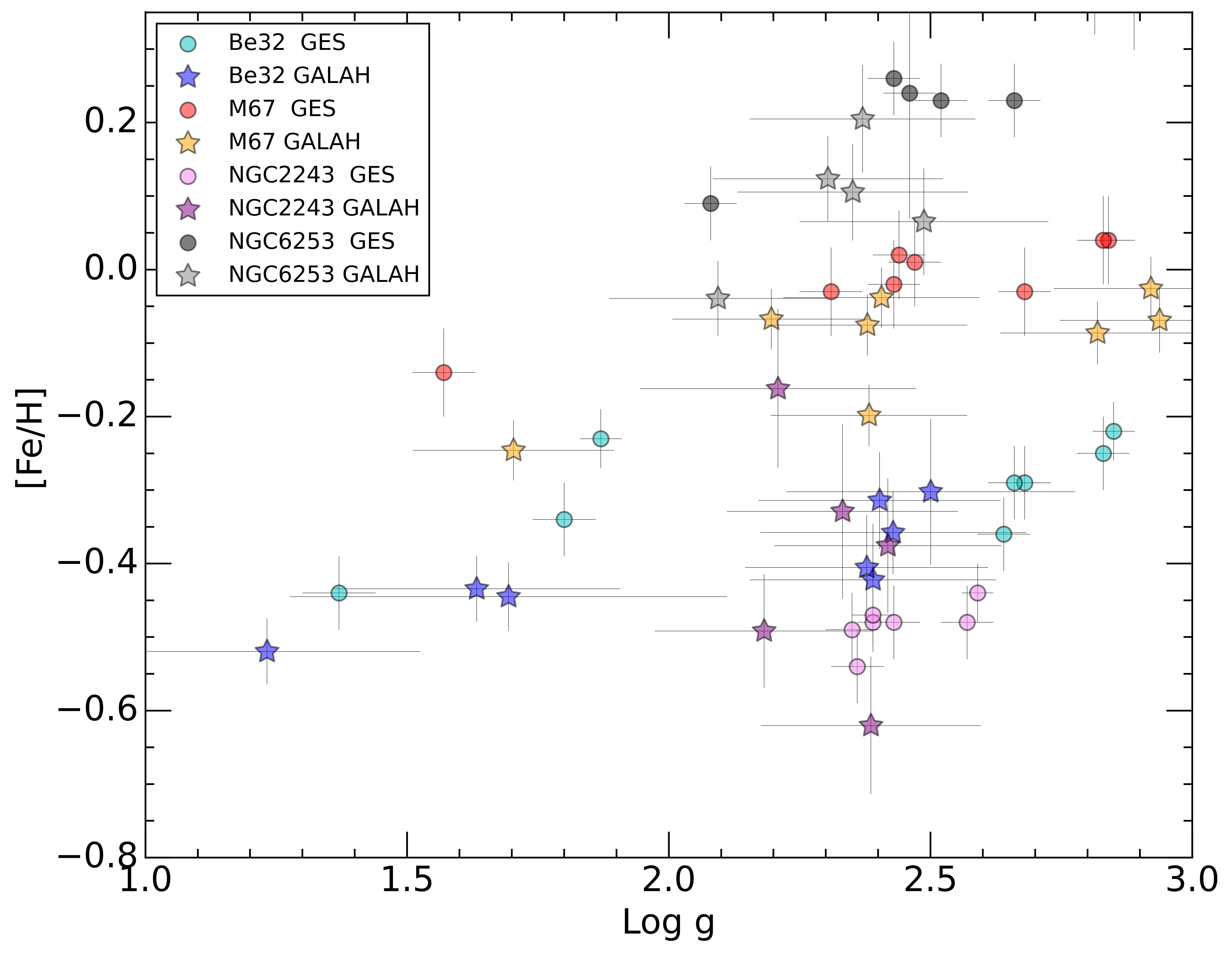}
%\resizebox{\hsize}{!}{\includegraphics{FIGURE/USED/cool_stars_ges_galah.png}}
  \caption{Metallicity vs. gravity of a sample of stars in clusters in common between {\em Gaia}-ESO and GALAH DR3. [Fe/H] and $\log g$ from GALAH DR3 and from {\em Gaia}-ESO are shown with different symbols, as indicated in the legend.      } 
  \label{Fig:galah}
\end{figure}

\begin{figure}
\includegraphics[scale=0.3]{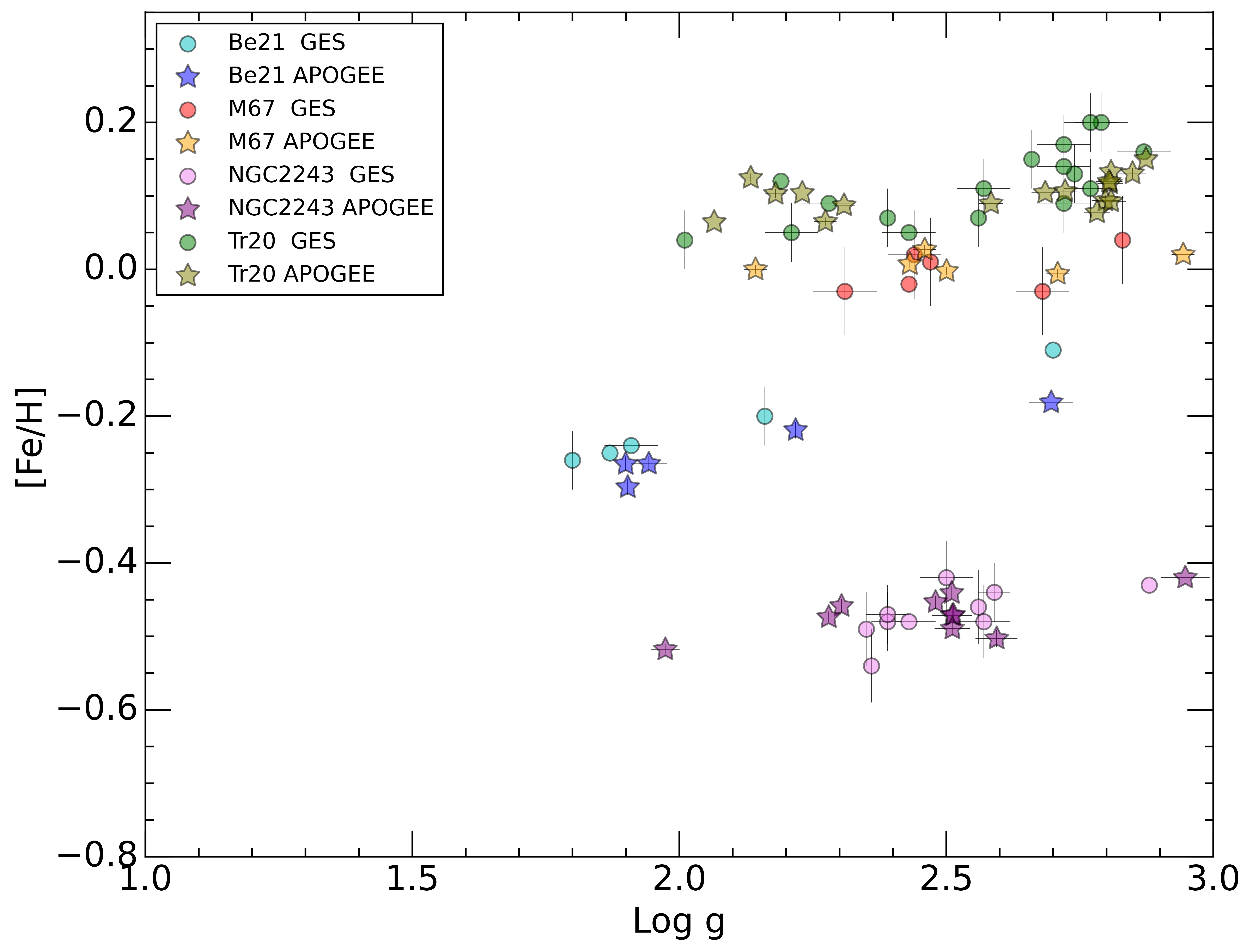}
%  \resizebox[scale=0.55]{!}{\includegraphics{FIGURE/USED/cool_stars_ges_apogee.png}}
  \caption{Metallicity vs. gravity of a  sample of stars in clusters in common between {\em Gaia}-ESO and APOGEE DR17. [Fe/H] and $\log g$ from APOGEE DR17 and from {\em Gaia}-ESO are shown with different symbols, as indicated in the legend.  } 
  \label{Fig:apogee}
\end{figure}

\begin{figure}
\includegraphics[scale=0.3]{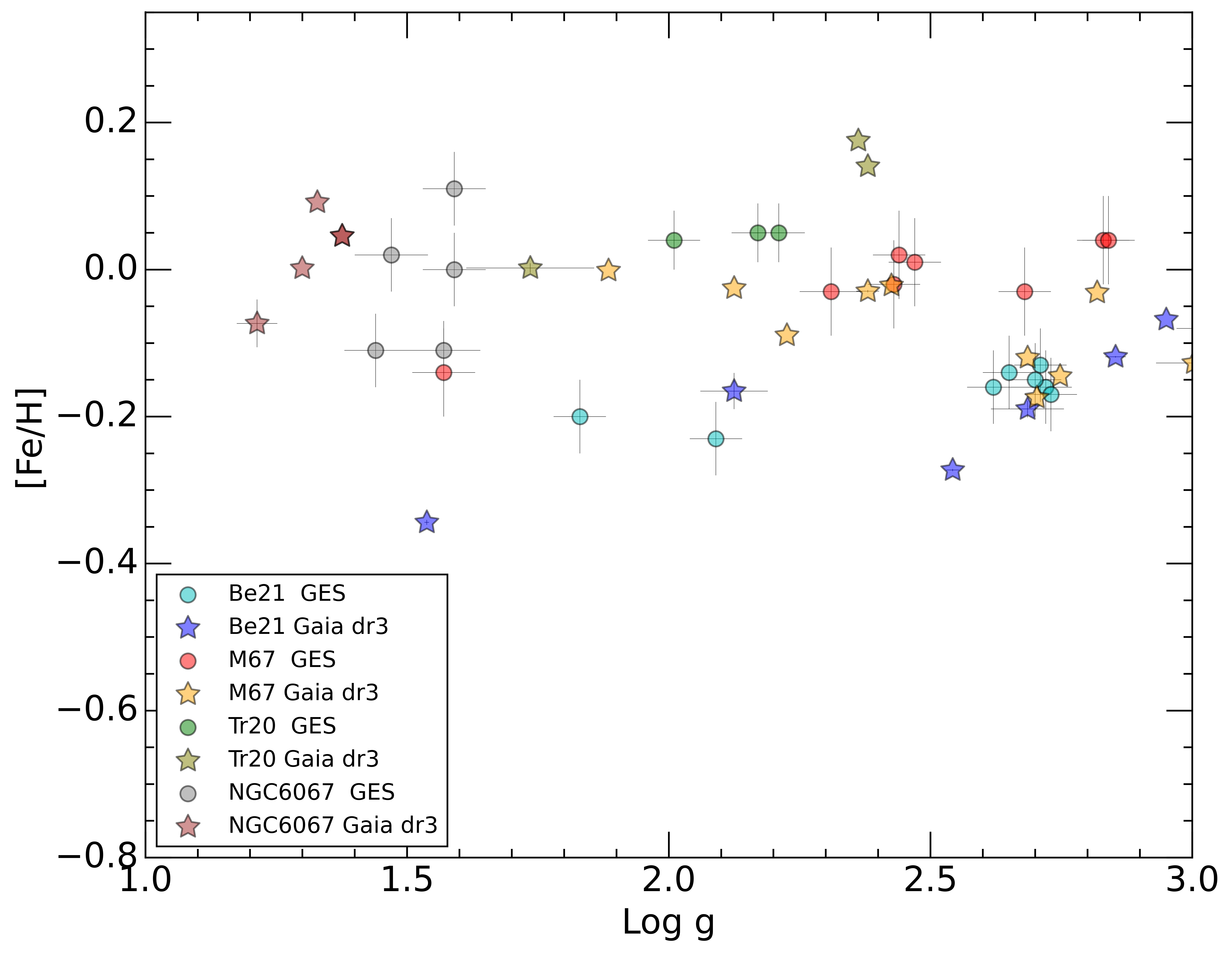}
%  \resizebox[scale=0.55]{!}{\includegraphics{FIGURE/USED/cool_stars_ges_gaia.png}}
  \caption{Metallicity vs. gravity of a  sample of stars in clusters in common between {\em Gaia}-ESO and {\em Gaia} RVS \citep{gaia22_parameters}. [Fe/H] and $\log g$ from {\em Gaia} and from {\em Gaia}-ESO are shown with different symbols, as indicated in the legend.  } 
  \label{Fig:gaia}
\end{figure}

In this section, we show the metallicity as a function of $\log g$ for a sample of members of open clusters in common between {\em Gaia}-ESO and GALAH DR3 and between {\em Gaia}-ESO and APOGEE DR17. In Fig.~\ref{Fig:galah} we present the results of GALAH and in Fig.~\ref{Fig:apogee} the results of APOGEE DR17. In both figures, a trend of decreasing metallicity as gravity declines is appreciable. In Fig.~\ref{Fig:gaia} we plot the results of {\em Gaia} RVS \citep{gaia22_parameters} for stars in common with {\em Gaia}-ESO.

\subsection{Individual element gradients: abundances and parameters of the WLS fits}
In this section, we provide tables with the Solar and M67 abundances (Table~\ref{solarabundance}), with coefficients of the WLS fitting for the two radial regions (Tables~\ref{tab:elh_rgc_coefficients} and \ref{tab:elfe_rgc_coefficients}) and for the three age bins (Tables~\ref{tab:elh_age_coefficients} and \ref{tab:elfe_age_coefficients}).

The average [El/H] and [El/Fe] of our cluster sample are  for the $\alpha$ elements in  Table~\ref{tab:average_alpha}, for the odd elements in Table~\ref{tab:average_odd}, for the iron-peak elements in Table~\ref{tab:average_iron}, for the $s$-process elements in Table~\ref{tab:average_s}, and finally, for the $r$-process elements in Table~\ref{tab:average_r}. In Table~\ref{tab:anumber} we show the number of member stars whose abundances have been used in each cluster per element. 
%The radial [El/H] and [El/Fe] gradients for the 24 considered elements are shown in Figs.~\ref{Fig:gradelemenetsh} and \ref{Fig:gradelemenetsfe} for the global samples, and in Figs.~\ref{Fig:gradelemenetshage} and \ref{Fig:gradelemenetsfeage} for clusters divided in the three age bins. 

\begin{table*}
\caption{{\sc iDR6} solar and M67 abundances.}              % title of Table
\label{solarabundance}      % is used to refer this table in the text
\centering                                      % used for centering table
\begin{tabular}{l c c c c c}          % centered columns (4 columns)
\hline\hline                        % inserts double horizontal lines
A(El) & Sun ({\sc iDR6}) & Sun & M67 ({\sc iDR6}) & M67 ({\sc iDR6}) & M67 ({\sc iDR6}) \\ 
& &  \citet{Grevesse_2007} & & (giants) & (dwarfs) \\ % table heading
\hline                                   % inserts single horizontal line
   $\ion{O}{i}$ &  8.66 $\pm 0.05 $&  8.66 $\pm 0.05$ & 8.74 $\pm$0.06  &  8.72 $\pm$0.06   & 8.80 $\pm$0.01 \\
   $\ion{Mg}{i}$  & 7.51 $\pm 0.02$& 7.53 $\pm 0.09$ & 7.50 $\pm $0.05  & 7.53 $\pm$0.03  & 7.48 $\pm$0.05 \\
   $\ion{Si}{i}$  & 7.44 $\pm 0.02$& 7.51 $\pm 0.04$ & 7.45 $\pm $0.06  & 7.50 $\pm$0.05 & 7.42 $\pm$0.06 \\
   $\ion{Ca}{i}$  & 6.32 $\pm 0.02$& 6.31 $\pm 0.04$ & 6.29 $\pm $0.06  & 6.25 $\pm$0.05 & 6.31 $\pm$0.06 \\
   $\ion{Ti}{i}$  & 4.89 $\pm 0.02$& 4.90 $\pm 0.06 $ & 4.87 $\pm$0.07  & 4.87 $\pm$0.07  & 4.87 $\pm$0.07 \\
   \hline 
   $\ion{Na}{i}$  &  6.18 $\pm 0.02 $&  6.17 $\pm0.04 $ & 6.22 $\pm $0.06  &  6.25 $\pm$0.06   & 6.20 $\pm$0.05 \\
\end{tabular}
\label{solarabundance}
\end{table*}

\newpage

\begin{table*}
\caption{\label{clusters_elfe} Average [$\alpha$/H] and [$\alpha$/Fe] for our sample of open clusters. The age (Gyr) and $R_\mathrm{GC}$ (kpc) are from \citet{CG20} and [Fe/H] from \citet{randich22}.}
\centering
% used for centering table
\scalebox{0.79}{
\begin{tabular}{lrrrrrrrrrrrrrrrr}
\hline\hline
GES\_FLD & [Fe/H] &  Age (Gyr) & R$_{\rm GC}$ (kpc) & $[\ion{O}{i}/$H] & $[\ion{Mg}{i}/$H] & $[\ion{Si}{i}/$H] &  $[\ion{Ca}{i}/$H]  & $[\ion{Ti}{i}/$H] & $[\ion{O}{i}/$Fe] & $[\ion{Mg}{i}/$Fe]  &  $[\ion{Si}{i}/$Fe] & $[\ion{Ca}{i}/$Fe] & $[\ion{Ti}{i}/$Fe]  \\
\hline
  Blanco1 & -0.03 & 0.1 & 8.3 &  & -0.01 & -0.03 & 0.01 & -0.04 &  & 0.01 & 0.01 & 0.03 & 0.0\\
  Br20 & -0.38 & 4.79 & 16.32 &  & -0.3 & -0.29 & -0.28 & -0.23 &  & 0.08 & 0.09 & 0.09 & 0.15\\
  Br21 & -0.21 & 2.14 & 14.73 &  & -0.17 & -0.17 & -0.26 & -0.26 &  & 0.03 & 0.04 & -0.05 & -0.05\\
  Br22 & -0.26 & 2.45 & 14.29 &  & -0.32 & -0.26 & -0.22 & -0.15 &  & -0.01 & 0.01 & 0.07 & 0.14\\
  Br25 & -0.25 & 2.45 & 13.81 &  & -0.29 & -0.26 & -0.3 & -0.28 &  & 0.01 & 0.05 & 0.04 & 0.03\\
  Br29 & -0.36 & 3.09 & 20.58 &  & -0.32 & -0.35 & -0.21 & -0.18 &  & 0.07 & 0.04 & 0.18 & 0.21\\
  Br30 & -0.13 & 0.3 & 13.25 & -0.12 & -0.18 & -0.14 & -0.14 & -0.32 & 0.05 & -0.04 & 0.01 & 0.0 & -0.18\\
\hline
\end{tabular}
\label{tab:average_alpha}}
\end{table*}

\begin{table*}
\caption{\label{clusters_elfe} Average [odd/H] and [odd/Fe] for our sample of open clusters. The age (Gyr) and $R_\mathrm{GC}$ (kpc) are from \citet{CG20} and [Fe/H] from \citet{randich22}.}
\centering
% used for centering table
\scalebox{0.87}{
\begin{tabular}{lrrrrrrrrrrrrrrrr}
\hline\hline
GES\_FLD & [Fe/H] &  Age (Gyr) & R$_{\rm GC}$ (kpc) & $[\ion{Na}{i}/$H] & $[\ion{Al}{i}/$H] & $[\ion{Cu}{i}/$H] &  $[\ion{Zn}{i}/$H] & $[\ion{Na}{i}/$Fe] & $[\ion{Al}{i}/$Fe]  &  $[\ion{Cu}{i}/$Fe] & $[\ion{Zn}{i}/$Fe]  \\
\hline
  Blanco1 & -0.03 & 0.1 & 8.3 & -0.09 & -0.1 & -0.21 & -0.06 & -0.06 & -0.07 & -0.19 & -0.02\\
  Br20 & -0.38 & 4.79 & 16.32 & -0.28 & -0.23 &  & -0.33 & 0.1 & 0.15 &  & 0.04\\
  Br21 & -0.21 & 2.14 & 14.73 & -0.28 & -0.29 & -0.62 & -0.13 & -0.04 & -0.09 & -0.42 & 0.07\\
  Br22 & -0.26 & 2.45 & 14.29 & -0.19 & -0.24 & -0.44 & -0.08 & 0.11 & 0.03 & -0.18 & 0.18\\
  Br25 & -0.25 & 2.45 & 13.81 & -0.26 & -0.3 & -0.24 & -0.26 & 0.05 & 0.0 & -0.08 & 0.05\\
  Br29 & -0.36 & 3.09 & 20.58 & -0.23 & -0.24 & -0.46 & -0.28 & 0.16 & 0.17 & -0.07 & 0.11\\
  Br30 & -0.13 & 0.3 & 13.25 & -0.09 & -0.24 &  & -0.15 & 0.06 & -0.09 &  & -0.0\\
\hline
\end{tabular}
\label{tab:average_odd}}
\end{table*}

\begin{table*}
\caption{\label{clusters_elfe} Average [iron peak/H] and [iron peak/Fe] for our sample of open clusters. The age (Gyr) and $R_\mathrm{GC}$ (kpc) are from \citet{CG20} and [Fe/H] from \citet{randich22}.}
\centering
% used for centering table
\scalebox{0.67}{
\begin{tabular}{lrrrrrrrrrrrrrrrr}
\hline\hline
GES\_FLD & [Fe/H] &  Age (Gyr) & R$_{\rm GC}$ (kpc) & $[\ion{Sc}{ii}/$H] & $[\ion{V}{i}/$H] & $[\ion{Cr}{i}/$H] &  $[\ion{Mn}{i}/$H] &  $[\ion{Co}{i}/$H] &  $[\ion{Ni}{i}/$H] & $[\ion{Sc}{ii}/$Fe] & $[\ion{V}{i}/$Fe]  &  $[\ion{Cr}{i}/$Fe] & $[\ion{Mn}{i}/$Fe] & $[\ion{Co}{i}/$Fe] & $[\ion{Ni}{i}/$Fe]\\
\hline

  Blanco1 & -0.03 & 0.1 & 8.3 & -0.14 & 0.01 & -0.02 & -0.04 & -0.16 & -0.13 & -0.1 & 0.05 & 0.02 & -0.01 & -0.12 & -0.09\\
  Br20 & -0.38 & 4.79 & 16.32 & -0.25 & -0.31 & -0.34 & -0.47 & -0.3 & -0.35 & 0.12 & 0.06 & 0.04 & -0.09 & 0.08 & 0.02\\
  Br21 & -0.21 & 2.14 & 14.73 & -0.12 & -0.26 & -0.24 & -0.32 & -0.27 & -0.28 & 0.09 & -0.06 & -0.03 & -0.08 & -0.06 & -0.07\\
  Br22 & -0.26 & 2.45 & 14.29 & -0.09 & -0.22 & -0.03 & -0.18 & -0.25 & -0.31 & 0.19 & 0.05 & 0.21 & 0.1 & 0.02 & -0.03\\
  Br25 & -0.25 & 2.45 & 13.81 & -0.09 & -0.32 & -0.27 & -0.25 & -0.24 & -0.45 & 0.18 & -0.01 & 0.06 & 0.05 & 0.07 & -0.1\\
  Br29 & -0.36 & 3.09 & 20.58 & -0.18 & -0.37 & -0.27 & -0.21 & -0.23 & -0.31 & 0.21 & 0.02 & 0.14 & 0.17 & 0.16 & 0.07\\
  Br30 & -0.13 & 0.3 & 13.25 & -0.14 & -0.22 & -0.15 & -0.33 & -0.25 & -0.29 & 0.0 & -0.08 & -0.0 & -0.18 & -0.1 & -0.15\\
\hline
\end{tabular}
\label{tab:average_iron}}
\end{table*}

\begin{table*}
\caption{\label{clusters_elfe} Average [$s$/H] and [$s$/Fe] for our sample of open clusters. The age (Gyr) and $R_\mathrm{GC}$ (kpc) are from \citet{CG20} and [Fe/H] from \citet{randich22}.}
\centering
% used for centering table
\scalebox{0.77}{
\begin{tabular}{lrrrrrrrrrrrrrrrr}
\hline\hline
GES\_FLD & [Fe/H] &  Age (Gyr) & R$_{\rm GC}$ (kpc) & $[\ion{Y}{ii}/$H] & $[\ion{Zr}{i}/$H] & $[\ion{Ba}{ii}/$H] &  $[\ion{La}{ii}/$H] &  $[\ion{Ce}{ii}/$H] & $[\ion{Y}{ii}/$Fe] & $[\ion{Zr}{i}/$Fe]  &  $[\ion{Ba}{ii}/$Fe] & $[\ion{La}{ii}/$Fe] & $[\ion{Ce}{ii}/$Fe] \\
\hline
  Blanco1 & -0.03 & 0.1 & 8.3 & 0.01 & -0.04 & 0.05 & 0.1 & -0.01 & 0.03 & 0.02 & 0.09 & 0.11 & 0.04\\
  Br20 & -0.38 & 4.79 & 16.32 & -0.34 & -0.35 & -0.32 & -0.35 & -0.34 & 0.04 & 0.02 & 0.05 & 0.08 & 0.04\\
  Br21 & -0.21 & 2.14 & 14.73 & -0.08 & -0.22 & 0.05 &  & -0.03 & 0.13 & 0.01 & 0.29 &  & 0.19\\
  Br22 & -0.26 & 2.45 & 14.29 & -0.26 & -0.33 & 0.04 &  & -0.27 & 0.01 & -0.03 & 0.28 &  & 0.11\\
  Br25 & -0.25 & 2.45 & 13.81 & -0.22 & -0.26 & -0.27 & -0.29 & -0.3 & 0.08 & 0.04 & 0.06 & 0.07 & 0.03\\
  Br29 & -0.36 & 3.09 & 20.58 & -0.1 &  & -0.37 &  &  & 0.29 &  & -0.01 &  & \\
  Br30 & -0.13 & 0.3 & 13.25 & -0.06 & -0.09 & 0.12 & 0.11 & 0.06 & 0.09 & 0.08 & 0.26 & 0.26 & 0.21\\
\hline
\end{tabular}
\label{tab:average_s}}
\end{table*}

\begin{table*}
\caption{\label{clusters_elfe} Average [$r$/H] and [$r$/Fe] for our sample of open clusters. The age (Gyr) and $R_\mathrm{GC}$ (kpc) are from \citet{CG20} and [Fe/H] from \citet{randich22}.}
\centering
% used for centering table
\scalebox{0.87}{
\begin{tabular}{lrrrrrrrrrrrrrrrr}
\hline\hline
GES\_FLD & [Fe/H] &  Age (Gyr) & R$_{\rm GC}$ (kpc) & $[\ion{Mo}{i}/$H] & $[\ion{Pr}{ii}/$H] & $[\ion{Nd}{ii}/$H] &  $[\ion{Eu}{ii}/$H] &  $[\ion{Mo}{i}/$Fe] & $[\ion{Pr}{ii}/$Fe]  &  $[\ion{Nd}{ii}/$Fe] & $[\ion{Eu}{ii}/$Fe]] \\
\hline
  Blanco1 & -0.03 & 0.1 & 8.3 &  & -0.09 & 0.14 & 0.06 &  & -0.06 & 0.18 & 0.07\\
  Br20 & -0.38 & 4.79 & 16.32 & -0.16 & -0.31 & -0.15 & -0.16 & 0.22 & 0.07 & 0.23 & 0.22\\
  Br21 & -0.21 & 2.14 & 14.73 & -0.23 & -0.12 & 0.14 & -0.07 & -0.0 & 0.11 & 0.35 & 0.16\\
  Br22 & -0.26 & 2.45 & 14.29 & 0.11 & -0.24 & 0.02 & 0.05 & 0.41 & 0.14 & 0.28 & 0.29\\
  Br25 & -0.25 & 2.45 & 13.81 & 0.03 & -0.16 & 0.02 & -0.14 & 0.34 & 0.18 & 0.32 & 0.18\\
  Br29 & -0.36 & 3.09 & 20.58 & 0.03 &  & -0.02 & 0.18 & 0.47 &  & 0.39 & 0.58\\
  Br30 & -0.13 & 0.3 & 13.25 & -0.18 & 0.01 & 0.12 & -0.0 & -0.01 & 0.16 & 0.27 & 0.15\\
\hline
\end{tabular}
\label{tab:average_r}}
\end{table*}

\begin{table*}
\caption{Number of member stars in each cluster and for each element used in this work. }
\centering
% used for centering table
\scalebox{0.7}{
%\hline
\begin{tabular}{lrrrrrrrrrrrrrrrrrrrrrrrr}
\hline \hline
GES\_FLD & OI & NaI & MgI & AlI & SiI & CaI & ScII & TiI & VI & CrI & MnI & CoI & NiI & CuI & ZnI & YII & ZrI & MoI & BaII & LaII & CeII &
PrII & NdII & EuII \\
\hline 
  Blanco1 & 0 & 9 & 14 & 11 & 14 & 14 & 9 & 13 & 14 & 13 & 14 & 11 & 12 & 8 & 13 & 11 & 2 & 0 & 10 & 3 & 3 & 3 & 6 & 2\\
  Br20 & 0 & 2 & 2 & 2 & 2 & 2 & 2 & 2 & 2 & 2 & 2 & 2 & 2 & 0 & 2 & 2 & 2 & 2 & 2 & 1 & 2 & 2 & 2 & 2\\
  Br21 & 0 & 11 & 12 & 12 & 10 & 12 & 12 & 12 & 12 & 10 & 12 & 12 & 12 & 3 & 11 & 7 & 6 & 7 & 10 & 0 & 7 & 7 & 11 & 8\\
  Br22 & 0 & 6 & 3 & 4 & 7 & 6 & 7 & 6 & 7 & 4 & 6 & 7 & 7 & 6 & 6 & 4 & 5 & 4 & 4 & 0 & 1 & 1 & 3 & 3\\
  Br25 & 0 & 6 & 6 & 6 & 6 & 5 & 5 & 6 & 5 & 5 & 6 & 6 & 4 & 1 & 6 & 3 & 5 & 4 & 5 & 4 & 4 & 3 & 4 & 3\\
  Br29 & 0 & 4 & 3 & 2 & 3 & 3 & 3 & 4 & 4 & 3 & 3 & 3 & 2 & 3 & 4 & 4 & 0 & 1 & 2 & 0 & 0 & 0 & 2 & 1\\
  Br30 & 1 & 3 & 3 & 3 & 3 & 3 & 3 & 3 & 3 & 3 & 3 & 3 & 3 & 0 & 3 & 3 & 1 & 1 & 3 & 3 & 3 & 3 & 3 & 3\\
\hline
\end{tabular}
\label{tab:anumber}}
\end{table*}

\begin{table*}

\caption{WLS fitting coefficients of [El/H] = $m_{1}\cdot$R$_{\rm GC}$ + c for the open clusters in the two investigated galactocentric regions and in the whole distance range.}              % title of Table
\label{}      % is used to refer this table in the text
\centering  
% used for centering table
\scalebox{0.8}{
\begin{tabular}{lrcccccccr}          % centered columns (4 columns)
\hline\hline % inserts double horizontal lines
 & & R$_{\rm GC}\leq$11.2 kpc & & & R$_{\rm GC}$ $>$  11.2 kpc  & & & global &\\
\hline
 & $m_{1}$ & c & PCC & $m_{1}$ & c & PCC & $m_{1}$ & c & PCC \\    % table heading
\hline
   $[\ion{Fe}/\ion{H}]$ & $-$0.081$\pm$0.008 & 0.692$\pm$ 0.068 & $-$0.858 & $-$0.044$\pm$0.014 & 0.376$\pm$0.178 & $-$0.609 & $-$0.054$\pm$0.004 & 0.474$\pm$0.045 & $-$0.851 \\

\hline 

  $[\ion{O}{i}/\ion{H}]$ & $-$0.051$\pm$0.009 & 0.416$\pm$0.069 & $-$0.755 &  &  & & $-$0.051$\pm$0.009 & 0.416$\pm$0.069 & $-$0.755\\
  $[\ion{Mg}{i}/\ion{H}]$ & $-$0.096$\pm$0.008 & 0.796$\pm$0.065 & $-$0.900 & $-$0.041$\pm$0.012 & 0.386$\pm$0.158 & $-$0.645 & $-$0.058$\pm$0.006 & 0.498$\pm$0.060 & $-$0.780\\
  $[\ion{Si}{i}/\ion{H}]$ & $-$0.088$\pm$0.011 & 0.725$\pm$0.089 & $-$0.803 & $-$0.023$\pm$0.004 & 0.118$\pm$0.063 & $-$0.794 & $-$0.039$\pm$0.004 & 0.325$\pm$0.045 & $-$0.762\\
  $[\ion{Ca}{i}/\ion{H}]$ & $-$0.059$\pm$0.008 & 0.514$\pm$0.058 & $-$0.778 & $-$0.011$\pm$0.004 & $-$0.001$\pm$0.063 & $-$0.521 & $-$0.034$\pm$0.003 & 0.326$\pm$0.030 & $-$0.805\\
  $[\ion{Ti}{i}/\ion{H}]$ & $-$0.065$\pm$0.008 & 0.514$\pm$0.070 & $-$0.787 & $-$0.010$\pm$0.010 & $-$0.049$\pm$0.127 & $-$0.238 & $-$0.045$\pm$0.005 & 0.358$\pm$0.047 & $-$0.756\\
\hline
\end{tabular}}
\label{tab:elh_rgc_coefficients}
\end{table*}

\begin{table*}

\caption{WLS fitting coefficients of [El/Fe] = $m_{1}\cdot$R$_{\rm GC}$ + c for the open clusters in the two investigated galactocentric regions and in the whole distance range.}              
% title of Table
\label{}      % is used to refer this table in the text
\centering  
% used for centering table
\scalebox{0.8}{
\begin{tabular}{lrcccccccr}          % centered columns (4 columns)
\hline\hline % inserts double horizontal lines
 & & R$_{\rm GC}\leq$11.2 kpc & & & R$_{\rm GC}$ $>$  11.2 kpc  & & & global &\\
\hline
 & $m_{1}$ & c & PCC & $m_{1}$ & c & PCC & $m_{1}$ & c & PCC \\    % table heading

\hline 

   $[\ion{O}{i}/\ion{Fe}]$ & 0.048$\pm$ 0.010 & $-$0.402$\pm$ 0.078 & 0.690  &  &  &  & 0.048$\pm$0.009 & $-$0.401$\pm$0.067 & 0.729\\
   $[\ion{Mg}{i}/\ion{Fe}]$ & 0.011$\pm$0.006 & $-$0.110$\pm$0.056 & 0.262 & 0.008$\pm$0.006 & $-$0.076$\pm$0.085 & 0.287 & 0.009$\pm$0.003 & $-$0.097$\pm$0.029 & 0.390\\
   $[\ion{Si}{i}/\ion{Fe}]$ & 0.000$\pm$0.004 & $-$0.009$\pm$0.037 & 0.012 & 0.004$\pm$0.005 & $-$0.054$\pm$0.070 & 0.199 & 0.002$\pm$0.002 & $-$0.021$\pm$0.019 & 0.130\\
   $[\ion{Ca}{i}/\ion{Fe}]$ & 0.041$\pm$0.004 & $-$0.355$\pm$0.031 & 0.866 & 0.002$\pm$0.008 & $-$0.005$\pm$0.102 & 0.062 & 0.018$\pm$0.003 & $-$0.184$\pm$0.027 & 0.653\\
   $[\ion{Ti}{i}/\ion{Fe}]$ & 0.027$\pm$0.007 & $-$0.253$\pm$0.055 & 0.559 & 0.024$\pm$0.008 & $-$0.321$\pm$0.101 & 0.593 & 0.012$\pm$0.003 & $-$0.135$\pm$0.031 & 0.433\\

\hline

  $[\ion{Na}{i}/\ion{Fe}]$ & $-$0.012$\pm$0.004 & 0.136$\pm$0.038 & $-$0.397 & 0.010$\pm$0.005 & $-$0.072$\pm$0.078 & 0.392 & 0.003$\pm$0.002 & 0.014$\pm$0.023 & 0.212\\
\hline
\end{tabular}}
\label{tab:elfe_rgc_coefficients}
\end{table*}

\begin{table*}

\caption{WLS fitting coefficients of [El/H] = $m_{1}\cdot$R$_{\rm GC}$ + c for the open clusters in the three investigated age bins.}              % title of Table
\label{}      % is used to refer this table in the text
\centering  
% used for centering table
\scalebox{0.8}{
\begin{tabular}{lrcccccccr}          % centered columns (4 columns)
\hline\hline % inserts double horizontal lines
 & & age (Gyr) < 1  & & &  1 $\leq$ age (Gyr) $\leq$ 3   & & & age (Gyr) > 3 &\\
\hline
 & $m_{1}$ & c & PCC & $m_{1}$ & c & PCC & $m_{1}$ & c & PCC \\    % table heading
\hline
   $[\ion{Fe}/\ion{H}]$ & -0.038$\pm$0.004 & 0.313$\pm$0.042 & $-$0.888 & $-$0.063$\pm$0.006 & 0.584$\pm$0.064 & $-$0.923 & $-$0.084$\pm$0.019 & 0.753$\pm$0.186 & $-$0.788\\
\hline 
   $[\ion{O}{i}/\ion{H}]$ & $-$0.037$\pm$0.014 & 0.332$\pm$0.105 & $-$0.597 & $-$0.055$\pm$0.027 & 0.444$\pm$0.183 & $-$0.755 & $-$0.040$\pm$0.018 & 0.298$\pm$0.162 & $-$0.653\\
   $[\ion{Mg}{i}/\ion{H}]$ & $-$0.069$\pm$0.013 & 0.539$\pm$0.093 & $-$0.767 & $-$0.044$\pm$0.007 & 0.390$\pm$0.079 & $-$0.797 & $-$0.082$\pm$0.020 & 0.713$\pm$0.208 & $-$0.771\\
   $[\ion{Si}{i}/\ion{H}]$ & $-$0.057$\pm$0.009 & 0.429$\pm$0.074 & $-$0.797 & $-$0.051$\pm$0.006 & 0.484$\pm$0.071 & $-$0.875 & $-$0.041$\pm$0.010 & 0.393$\pm$0.131 & $-$0.754\\
   $[\ion{Ca}{i}/\ion{H}]$ & $-$0.045$\pm$0.005 & 0.410$\pm$0.037 & $-$0.885 & $-$0.042$\pm$0.005 & 0.414$\pm$0.045 & $-$0.890 & $-$0.017$\pm$0.008 & 0.090$\pm$0.101 & $-$0.540\\
   $[\ion{Ti}{i}/\ion{H}]$ & $-$0.052$\pm$0.004 & 0.384$\pm$0.041 & $-$0.934 & $-$0.036$\pm$0.005 & 0.308$\pm$0.046 & $-$0.834 & $-$0.049$\pm$0.018 & 0.447$\pm$0.162 & $-$0.624\\
\hline
\end{tabular}}
\label{tab:elh_age_coefficients}
\end{table*}

\begin{table*}

\caption{WLS fitting coefficients of [El/Fe] = $m_{1}\cdot$R$_{\rm GC}$ + c for the open clusters in the three investigated age bins.}              % title of Table
\label{}      % is used to refer this table in the text
\centering  
% used for centering table
\scalebox{0.8}{
\begin{tabular}{lrcccccccr}          % centered columns (4 columns)
\hline\hline % inserts double horizontal lines
 & & age (Gyr) < 1  & & &  1 $\leq$ age (Gyr) $\leq$ 3   & & & age (Gyr) > 3 &\\
\hline
 & $m_{1}$ & c & PCC & $m_{1}$ & c & PCC & $m_{1}$ & c & PCC \\    % table heading

\hline 
  $[\ion{O}{i}/\ion{Fe}]$ & 0.014$\pm$0.012 & $-$0.116$\pm$0.094 & 0.304 & 0.022$\pm$0.010 & $-$0.272$\pm$0.068 & 0.777 & 0.093$\pm$0.018 & $-$0.864$\pm$0.179 & 0.892\\
  $[\ion{Mg}{i}/\ion{Fe}]$ & $-$0.014$\pm$0.011 & 0.091$\pm$0.091 & $-$0.272 & 0.010$\pm$0.003 & $-$0.119$\pm$0.027 & 0.640 & 0.014$\pm$0.004 & $-$0.124$\pm$0.044 & 0.723\\
  $[\ion{Si}{i}/\ion{Fe}]$ & $-$0.009$\pm$0.006 & 0.055$\pm$0.051 & $-$0.303 & 0.003$\pm$0.002 & $-$0.036$\pm$0.020 & 0.354 & 0.004$\pm$0.003 & $-$0.026$\pm$0.029 & 0.374\\
  $[\ion{Ca}{i}/\ion{Fe}]$ & 0.006$\pm$0.004 & $-$0.045$\pm$0.036 & 0.310 & 0.020$\pm$0.003 & $-$0.230$\pm$0.032 & 0.787 & 0.026$\pm$0.007 & $-$0.222$\pm$0.067 & 0.759\\
  $[\ion{Ti}{i}/\ion{Fe}]$ & $-$0.001$\pm$0.005 & $-$0.025$\pm$0.049 & $-$0.055 & 0.021$\pm$0.003 & $-$0.253$\pm$0.033 & 0.806 & 0.016$\pm$0.006 & $-$0.129$\pm$0.050 & 0.632\\

\hline
  $[\ion{Na}{i}/\ion{Fe}]$ & 0.001$\pm$0.005 & 0.053$\pm$0.047 & 0.051 & 0.002$\pm$0.003 & $-$0.011$\pm$0.030 & 0.195 & 0.006$\pm$0.003 & $-$0.007$\pm$0.044 & 0.443\\
\hline
\end{tabular}}
\label{tab:elfe_age_coefficients}
\end{table*}

\end{appendix}

\end{document}